\documentclass{elsarticle}
\usepackage{amsmath,amssymb,amsfonts}
\usepackage{graphicx}
\usepackage{xcolor}
\journal{Journal of Sound and Vibration}

\begin{document}
\title{Investigation upon the performance of piezoelectric energy harvester with elastic extensions}

\author[1]{Maoying Zhou \corref{cor1} }
\ead{myzhou@hdu.edu.cn}
\author[2,3]{Yang Fu }
\ead{yangfu@zust.edu.cn}
\author[1]{Huawei Ji}
\ead{jhw76@163.com}

\cortext[cor1]{Corresponding author}

\address[1]{School of Mechanical Engineering, Hangzhou Dianzi University, Hangzhou, China}
\address[2]{School of Mechanical and Automotive Engineering, Zhejiang University of Science and Technology, Hangzhou, China}
\address[3]{The State Key Laboratory of Fluid Power and Mechatronic Systems, Zhejiang University, Hangzhou, China}

\begin{abstract}

Piezoelectric vibration energy harvesters have attracted much attention due to its potential to replace currently popular batteries and to provide an sustainable power sources. Many researchers have proposed ways to increase the performance of piezoelectric energy harvesters like bandwidth, working frequency and output performance. Here in this contribution, we propose the method of using elastic extensions to tune the performance of a piezoelectric energy harvester. Mathematical model of the proposed device is derived and analyzed. Numerical simulations are done to investigate the influences of the derived parameters, like length ratio $\lambda_l$, bending stiffness ratio $\lambda_B$, and line density ratio $\lambda_m$. Results show that the elastic extension does change the motion of the proposed device and help tune the performance of piezoelectric energy harvesters. 

\end{abstract}


\maketitle

\section{Introduction}

To address the urgent issues of global warming, environmental pollution, and depleting energy resources, researchers have been continuously exploring clean, renewable, and sustainable energy sources, including solar energy, bio-energy, nuclear energy and etc. \cite{zhou2018review} Extracting energy from ambient environment to power electronic devices, which is also called energy harvesting, has also been a hot research topic in the past few decades. Lots of energy harvesting devices have been put forward to convert ambiently available energy, such as vibration energy, thermal energy, solar energy, and so on, into electrical energy. \cite{dagdeviren2017energy,safaei2019review} Among all these devices, vibration energy harvesting devices have attracted the most attention due to the universal and versatile presence of vibration energy in engineering applications. \cite{liu2018comprehensive}

According to the underlying physical principles and materials used in the devices, vibration energy harvesting devices can be classified into several categories: electromagnetic energy harvesting devices \cite{cepnik2013review,tan2016review}, electrostatic energy harvesting devices \cite{khan2016state,suzuki2011recent}, piezoelectric energy harvesting devices \cite{liu2018comprehensive,sodano2004review}, and triboelectric energy harvesting devices \cite{fan2016flexible,wang2015progress}. Owing to their simplicity of structure and ease of fabrication, piezoelectric energy harvesting devices has been the focus of many researchers in the past twenty years. \cite{anton2007review,safaei2019review}

A classic piezoelectric energy harvester (CPEH) is composed of an elastic beam and some piezoelectric elements attached to the beam. Theoretical modelling, numerical computation, and experimental investigations have been conducted to analyze the performances of CPEHs. \cite{erturk2008distributed,erturk2009experimentally} Different methods have been put forward to optimize the performance of CPEHs by changing beam shape \cite{goldschmidtboeing2008characterization}, electrode coverage \cite{fu2018electrode} and configuration \cite{kim2015effect}, streamwise position of piezoelectric patches \cite{liao2012optimal,patel2011geometric}, and etc. Despite all these efforts, the performance of current piezoelectric energy harvesting devices are still far from meeting the requirement of engineering applications. They are typically confronted with limited bandwidth and poor energy conversion efficiency. To improve the performance of piezoelectric energy harvesting devices, many techniques, such as resonance tuning, motion amplification, nonlinearity introduction, and so on, are presented, investigated, and analyzed. \cite{yildirim2017review,tran2018ambient} These researches help us obtain an indepth understanding of the piezoelectric energy harvesting mechanisms and provide useful ways to improve the performance of piezoelectric energy harvesters. However, they do not take into account the possible effect of elastic extensions, which has been shown feasible in improving output performance of piezoelectric wind energy harvesters.\cite{zhou2019piezoelectric}

Here in this contribution, we focus on the influence of elastic extension upon the performance of piezoelectric energy harvesters. A piezoelectric energy harvester with attached elastic extension is put forward. Its mathematical model is derived based on Euler-Bernoulli beam model, and converted into a boundary value problem. Numerical simulations are conducted to investigate the influences of system parameters upon its performances. Results validate our proposition and provide insights into the improvement of piezoelectric energy harvesting.

\section{Modeling of the proposed system}

\begin{figure}[!htbp]
    \centering
    \includegraphics[width=0.8\textwidth]{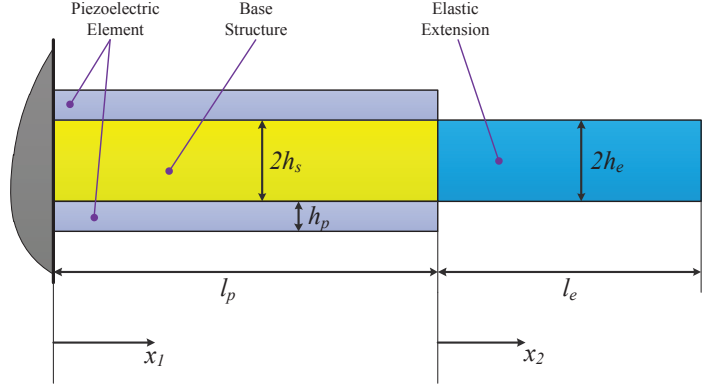}
    \caption{Schematic configuration of a piezoelectric energy harvester with elastic extension (PEHEE). }
    \label{fig:fig_beam_configuration}
\end{figure}

The proposed piezoelectric energy harvester with elastic extension (PEHEE) is similar to a CPEH, as shown in Figure~\ref{fig:fig_beam_configuration}. The PEHEE is made up of two parts: the primary beam and the extension beam. The primary beam is simply a CPEH, which is usually comprised of the base structure and the piezoelectric elements attached to it. Note that the primary beam is usually of a layered structure and there can be many layers of base structure as well as piezoelectric elements. Here in Figure~\ref{fig:fig_beam_configuration}, we shown the structure of a bimorph as an example. The base structure is of thickness $2h_s$ while the piezoelectric elements are of thickness $h_p$. The total length of the primary beam is $l_p$. The extension beam is simply made from certain elastic materials, like PET. Its thickness is $2h_e$ and its length is $l_e$. Both the spanwise widths of the primary beam and the extension beam are set to be identical to $b$.

Following the classical analyzing process of piezoelectric bimorph cantilever beams \cite{erturk2008distributed,erturk2009experimentally,park2003dynamics}, the constitutive equations for the primary beam are listed as:
\begin{equation}
    \left\{\begin{aligned}
        M_p(x_1,t) &= B_p \frac{\partial^2 w_1(x_1,t)}{\partial x_1^2} - e_p V_p (t) \\
        q_p(x_1,t) &= e_p \frac{\partial^2 w_1(x_1,t)}{\partial x_1^2} + \epsilon_p V_p (t),
    \end{aligned}\right.
    \label{eq:eq_primary_beam_consitutive_equation}
\end{equation}
where $M_p(x_1,t)$ is the bending moment at the cross section of position coordinate $x_1$ and $q_p(x_1,t)$ is the corresponding local line charge density on the electrode. $w_1(x_1,t)$ is the displacement function of the primary beam with $0 \leq x_1 \leq l_p$ and $V_p(t)$ is the voltage across the electrodes. Considering that the piezoelectric elements are arranged in a serial connection manner \cite{erturk2009experimentally}, the coefficients $B_p$, $e_p$, and $\epsilon_p$ are defined as
\begin{equation}
    \left\{\begin{aligned}
        B_p &= \frac{2}{3}b\left\{ Y_s h_s^3 + Y_p \left[ (h_s+h_p)^3 - h_s^3 \right] \right\}, \\
        e_p &= b e_{31}\left(h_s+\frac{1}{2}h_p\right), \\
        \epsilon_p &= \frac{b \epsilon_{33}^S}{2 h_p},
    \end{aligned}\right.
    \label{eq:eq_primary_beam_consitutive_equation_coefficients}
\end{equation}
in which $Y_p$ and $Y_s$ are the elastic constants of the piezoelectric element and the base structure, respectively, $e_{31}$ is the piezoelectric charge constant of the piezoelectric element, $\epsilon_{33}^S$ is the dielectric constant of the piezoelectric element.

Using Euler-Bernoulli assumptions \cite{Timoshenko1974Vibration}, dynamic equations of the primary beam are formulated as 
\begin{equation}
    B_p \frac{\partial^4 w_1(x_1,t)}{\partial x_1^4} + m_p \frac{\partial^2 w_1(x_1,t)}{\partial t^2} = 0
    \label{eq:eq_primary_beam_equilibrium_equation}
\end{equation}
where $m_p = 2 b(\rho_s h_s + \rho_p h_p)$ is the line mass density of the primary beam with $\rho_s$ and $\rho_p$ being the volumetric density of the base structure and the piezoelectric element, respectively. Principally piezoelectric energy harvester is regarded as a current source due to its dielectric property. Hence the charge accumulation $Q_p(t)$ on the electrode  is of special interest in our research. According to a simple integration process, it can be calculated as 
\begin{equation}
    Q_p(t) = \int_0^{l_p} q_p(x_1, t)\ d x_1\ = \ e_p \left.\left[ \frac{\partial w_1(x_1,t)}{\partial x_1} \right]\right|^{l_p}_0 + C_p V_p (t)
    \label{eq:eq_primary_beam_charge_accumulation}
\end{equation}
where $C_p = \epsilon_p l_p$ is the inherent capacitance of the piezoelectric layer in the primary beam. According to the Kirchhoff's law, the electric equilibrium equation is 
\begin{equation}
    \frac{d Q_p(t)}{dt}  + \frac{V_p(t)}{R_l} = 0
    \label{eq:eq_primary_beam_circuit_equation}
\end{equation}
where $R_l$ is the load resistance connected to the piezoelectric energy harvester.

For the extension beam, the governing equations are simply
\begin{equation}
    B_e \frac{\partial^4 w_2(x_2,t)}{\partial x_2^4} + m_e \frac{\partial^2 w_2(x_2,t)}{\partial t^2} = 0
    \label{eq:eq_extension_beam_equilibrium_equation}
\end{equation}
where $w_2(x_2,t)$ is the displacement of the extension beam at position $0\leq x_2 \leq l_e$, $B_e = \frac{2}{3} Y_e b h_e^3$ is the equivalent bending stiffness of the extension beam, $m_e = 2\rho_e h_e$ is the line mass density of the extension beam, $\rho_e$ is the volumetric mass density of the extension beam. In this process, $Y_e$ is the Young's modulus of the extension beam. As a result, the defining relations for the bending moment $M_e(x_2,t)$ at the position $x_2$ is
\begin{equation}
    M_e(x_2,t) = B_e \frac{\partial^2 w_2(x_2,t)}{\partial x_2^2}.
    \label{eq:eq_extension_beam_consitution_equation}
\end{equation}

The primary beam and the extension beam are connected to each other at one end. Therefore at the joint point $x_1 = l_p$ or $x_2 = 0$, we expect the continuity of displacement functions $w_1(x_1,t)$ and $w_2(x_2,t)$, bending moments $M_p(x_1,t)$ and $M_e(x_2,t)$, and shearing forces $\frac{\partial M_p(x_1,t)}{\partial x_1}$ and $\frac{\partial M_e(x_2,t)}{\partial x_2}$. That is to say, 
\begin{equation}
    \left\{\begin{aligned}
        w_1(l_p,t) &= w_2(0,t), \\
        \frac{\partial w_1(l_p,t)}{\partial x_1} &= \frac{\partial w_2(0,t)}{\partial x_2}, \\
        B_p \frac{\partial^2 w_1(l_p,t)}{\partial x_1^2} - e_p V_p(t) &= B_e \frac{\partial^2 w_2(0,t)}{\partial x_2^2}, \\
        B_p \frac{\partial^3 w_1(l_p,t)}{\partial x_1^3} &= B_e \frac{\partial^3 w_2(0,t)}{\partial x_2^3}.
    \end{aligned}\right.
    \label{eq:eq_connection_point_continuity}
\end{equation}
The left end of the primary beam is connected to the vibration sources whose vibration can be described as a function $w_b(t)$. To make it easier, we assume that the base excitation $w_b(t)$ is the only input to the system and $w_b(t)$ is strictly in the direction of $w_1(x_1,t)$ and $w_2(x_2,t)$ so that no rotation is present at the left end. As a result, we have 
\begin{equation}
    w_1(0,t) = w_b(t), \quad \frac{\partial w_1(0,t)}{\partial x_1} = 0.
    \label{eq:eq_left_end_base_excitation}
\end{equation}
At the right end of the extension beam where $x_2 = l_e$, no external forces and moments are applied and we have
\begin{equation}
    \frac{\partial^2 w_2(l_e,t)}{\partial x_2^2} = 0, \quad \frac{\partial^3 w_2(l_e,t)}{\partial x_2^3} = 0.
    \label{eq:eq_right_end_free_action}
\end{equation}

With the governing equations (\ref{eq:eq_primary_beam_consitutive_equation},\ref{eq:eq_primary_beam_equilibrium_equation}), (\ref{eq:eq_primary_beam_consitutive_equation_coefficients}), (\ref{eq:eq_primary_beam_charge_accumulation}), (\ref{eq:eq_primary_beam_circuit_equation}), (\ref{eq:eq_extension_beam_equilibrium_equation}), and (\ref{eq:eq_extension_beam_consitution_equation}), and the boundary conditions (\ref{eq:eq_connection_point_continuity}), (\ref{eq:eq_left_end_base_excitation}), and (\ref{eq:eq_right_end_free_action}), we set up a boundary value problem for the base excitation problem for the PEHEE. Generally, the problem is solved using mode decomposition method \cite{erturk2009experimentally} or finite element method \cite{maurini2006numerical}, where the actual motion is expressed using a series of accurate eigenmodes or approximate eigenmodes. Here in this contribution, as we are interested in the steady state response of the PEHEE, the harmonic balance method \cite{wu1992generalized} is used. 

Usually we consider a base excitation $w_b(t)$ of single angular frequency $\sigma$ and amplitude $\xi_b$, which can be expressed by $w_b(t) = Re \left\{ \xi_b e^{j \sigma t} \right\}$ with $j = \sqrt{-1}$ being the imaginary unit. The amplitude $\xi_b$ is complex and contains the phase information of the base excitation. To make things easier, we set $\xi_b$ to be a real constant designated by the controller. Hence, according to the harmonic balance method, we can set the steady state response of the displacements $w_1(x_1,t)$ and $w_2(x_2,t)$ of the primary beam and the extension beam respectively as 
\begin{equation}
    w_1(x_1,t) = \tilde{w}_1(x_1)e^{j \sigma t},\quad w_2(x_2,t) = \tilde{w}_2(x_2)e^{j \sigma t},
\end{equation}
the steady state voltage response $V_p(t)$ and charge accumulation $Q_p(t)$ as
\begin{equation}
    V_p(t) = \tilde{V}_p e^{j \sigma t},\quad Q_p(t) = \tilde{Q}_p e^{j \sigma t},
\end{equation}
and the bending moment $M_p(x_1,t)$ and $M_e(x_2,t)$ as
\begin{equation}
    M_p(x_1,t) = \tilde{M}_p(x_1) e^{j \sigma t},\quad M_e(x_2,t) = \tilde{M}_e(x_2) e^{j \sigma t},
\end{equation}
where the symbols with a hat $\sim$ denote the complex modulus of the corresponding quantities, which contains the amplitude and phase information. As a result, the above system of governing equations for the PEHEE can be summarized as
\begin{equation}
    \left\{\begin{aligned}
        B_p \frac{\partial^4 \tilde{w}_1(x_1)}{\partial x_1^4} - m_p \sigma^2 \tilde{w}_1(x_1) &= 0, \\
        B_e \frac{\partial^4 \tilde{w}_2(x_2)}{\partial x_2^4} - m_e \sigma^2 \tilde{w}_2(x_2) &= 0, \\
        j \sigma \tilde{Q}_p + \frac{\tilde{V}_p}{R_l} &= 0,
    \end{aligned}\right.
    \label{eq:eq_balance_equations_original}
\end{equation}
\begin{equation}
    \left\{\begin{aligned}
        \tilde{M}_p(x_1) &= B_p \frac{\partial^2 \tilde{w}_1(x_1)}{\partial x_1^2} - e_p \tilde{V}_p, \\
        \tilde{Q}_p &= \ e_p \left.\left[ \frac{\partial \tilde{w}_1(x_1)}{\partial x_1} \right]\right|^{l_p}_0 + C_p \tilde{V}_p, \\
        \tilde{M}_e(x_2) &= B_e \frac{\partial^2 \tilde{w}_2(x_2)}{\partial x_2^2},
    \end{aligned}\right.
    \label{eq:eq_constitutive_equations_original}
\end{equation}
and the boundary conditions become
\begin{equation}
    \left\{\begin{aligned}
        \tilde{w}_1(0) = \xi_b , &\quad \frac{\partial \tilde{w}_1}{\partial x_1} (0) = 0, \\
        w_1(l_p,t) = w_2(0,t), &\quad \frac{\partial \tilde{w}_1(l_p)}{\partial x_1} = \frac{\partial \tilde{w}_2(0)}{\partial x_2}, \\
        B_p \frac{\partial^2 \tilde{w}_1(l_p)}{\partial x_1^2} - e_p \tilde{V}_p = B_e \frac{\partial^2 \tilde{w}_2(0)}{\partial x_2^2} , &\quad B_p \frac{\partial^3 \tilde{w}_1(l_p)}{\partial x_1^3} = B_e \frac{\partial^3 \tilde{w}_2(0)}{\partial x_2^3}, \\
        \frac{\partial^2 \tilde{w}_2(l_e)}{\partial x_2^2} = 0 , &\quad \frac{\partial^3 \tilde{w}_2(l_e)}{\partial x_2^3} = 0.
    \end{aligned}\right.
    \label{eq:eq_boundary_conditions_original}
\end{equation}

From the equations (\ref{eq:eq_balance_equations_original}), (\ref{eq:eq_constitutive_equations_original}), and (\ref{eq:eq_boundary_conditions_original}), we can eliminate the electrical quantities $\tilde{Q}_p$ and $\tilde{V}_p$ by incorporating them into the boundary conditions. Actually, from equations (\ref{eq:eq_balance_equations_original}) and (\ref{eq:eq_constitutive_equations_original}), we have
\begin{equation}
    \tilde{V}_p = \frac{j \sigma R_l e_p}{j \sigma R_l C_p + 1} \left.\left[ \frac{\partial \tilde{w}_1(x_1)}{\partial x_1} \right]\right|^{l_p}_0,
\end{equation}
which can actually be used to eliminate the term $\tilde{V}_p$ in the boundary conditions (\ref{eq:eq_boundary_conditions_original}). In the end, we can simplify the problem as a combination of the governing equations
\begin{equation}
    \left\{\begin{aligned}
        B_p \frac{\partial^4 \tilde{w}_1(x_1)}{\partial x_1^4} - m_p \sigma^2 \tilde{w}_1(x_1) &= 0, \\
        B_e \frac{\partial^4 \tilde{w}_2(x_2)}{\partial x_2^4} - m_e \sigma^2 \tilde{w}_2(x_2) &= 0, 
    \end{aligned}\right.
    \label{eq:eq_balance_equations_converted}
\end{equation}
and the boundary conditions
\begin{equation}
    \left\{\begin{aligned}
        \tilde{w}_1(0) = \xi_b , &\quad \frac{\partial \tilde{w}_1}{\partial x_1} (0) = 0, \\
        \tilde{w}_1(l_p) = \tilde{w}_2(0), &\quad \frac{\partial \tilde{w}_1(l_p)}{\partial x_1} = \frac{\partial \tilde{w}_2(0)}{\partial x_2}, \\
        B_p \frac{\partial^2 \tilde{w}_1(l_p)}{\partial x_1^2} + \frac{j \sigma R_l e_p^2}{j \sigma R_l C_p + 1}  \frac{\partial \tilde{w}_1(l_p)}{\partial x_1} = B_e \frac{\partial^2 \tilde{w}_2(0)}{\partial x_2^2} , &\quad B_p \frac{\partial^3 \tilde{w}_1(l_p)}{\partial x_1^3} = B_e \frac{\partial^3 \tilde{w}_2(0)}{\partial x_2^3}, \\
        \frac{\partial^2 \tilde{w}_2(l_e)}{\partial x_2^2} = 0 , &\quad \frac{\partial^3 \tilde{w}_2(l_e)}{\partial x_2^3} = 0,
    \end{aligned}\right.
    \label{eq:eq_boundary_conditions_converted}
\end{equation}
which actually manifests as a boundary value problem.

\section{Dimensionless Problem}
Defining the following dimensionless variables
\begin{equation}
    \tilde{w}_1 = \xi_b u_1,\quad \tilde{w}_2 = \xi_b u_2,\quad \tilde{x}_1 = l_p x,\quad \tilde{x}_2 = l_e x.,
    \label{eq:eq_non_dim_variables}
\end{equation}
The above listed boundary value problem is made dimensionless. Note that here we use one independent space variable $x$ to nondimensionalize two previously used variables $x_1$ and $x_2$. This comes from the fact that the variables $x_1$ and $x_2$ are not coupled with each other in the sense that the primary beam and the extension beam  do not overlap each other except for their joint point where $x_1 = l_p$ and $x_2 = 0$. Thus the two variables do not occur in the equations simultaneously except for the boundary conditions. As for the boundary conditions, the change of variables does not affect the values of the equations. In one word, the equation (\ref{eq:eq_non_dim_variables}) does not change the problem in essence. Hence, the above boundary value problem is further changed into the combination of the governing equations
\begin{equation}
    \left\{\begin{aligned}
        \frac{B_p}{l_p^4} u_1^{\prime\prime\prime\prime} - m_p \sigma^2 u_1 &= 0, \\
        \frac{B_e}{l_e^4} u_2^{\prime\prime\prime\prime} - m_e \sigma^2 u_2 &= 0, 
    \end{aligned}\right.
    \label{eq:eq_balance_equations_nondim}
\end{equation}
and the boundary conditions
\begin{equation}
    \left\{\begin{aligned}
        u_1(0) = 1 , &\quad u_1^\prime(0) = 0, \\
        u_1(1) = u_2(0), &\quad \frac{1}{l_p} u_1^\prime(1) = \frac{1}{l_e} u_2^\prime(0), \\
        \frac{B_p}{l_p^2} u_1^{\prime\prime}(1) + \frac{j \sigma R_l e_p^2}{j \sigma R_l C_p + 1} \frac{1}{l_p} u_1^{\prime}(1) = \frac{B_e}{l_e^2} u_2^{\prime\prime}(0) , &\quad \frac{B_p}{l_p^3} u_1^{\prime\prime\prime}(1) = \frac{B_e}{l_e^3} u_2^{\prime\prime\prime}(0), \\
        u_2^{\prime\prime}(1) = 0 , &\quad u_2^{\prime\prime\prime}(1) = 0,
    \end{aligned}\right.
    \label{eq:eq_boundary_conditions_nondim}
\end{equation}
in which the prime means the derivative with respect to $x$. The equations can again be organized in a more compact form
\begin{equation}
    \left\{\begin{aligned}
         u_1^{\prime\prime\prime\prime} - \nu^2 u_1 &= 0, \\
         u_2^{\prime\prime\prime\prime} - \nu^2 \lambda_m \lambda_l^4 / \lambda_B u_2 &= 0,
    \end{aligned}\right.
    \label{eq:eq_balance_equations_compact}
\end{equation}
\begin{equation}
    \left\{\begin{aligned}
        u_1(0) = 1 , &\quad u_1^\prime(0) = 0, \\
        u_1(1) = u_2(0), &\quad \lambda_l u_1^\prime(1) = u_2^\prime(0), \\
        u_1^{\prime\prime}(1) + \frac{ j \nu \beta }{ j \nu \beta + 1 } \alpha^2 u_1^{\prime}(1) = \lambda_B/ \lambda_l^2 u_2^{\prime\prime}(0) , &\quad u_1^{\prime\prime\prime}(1) = \lambda_B/ \lambda_l^3 u_2^{\prime\prime\prime}(0), \\
        u_2^{\prime\prime}(1) = 0 , &\quad u_2^{\prime\prime\prime}(1) = 0,
    \end{aligned}\right.
    \label{eq:eq_boundary_conditions_compact}
\end{equation}
where 
\begin{equation}
    \nu = \sigma \sqrt{ \frac{ m_p l_p^4 }{ B_p } },\quad \lambda_B = \frac{B_e}{B_p},\quad \lambda_m = \frac{m_e}{m_p},\quad \lambda_l = \frac{l_e}{l_p},
\end{equation}
\begin{equation}
    \beta = R_l C_p \sqrt{\frac{B_p}{m_p l_p^4}}, \quad \alpha = e_p \sqrt{\frac{l_p}{C_p B_p}}.
\end{equation}

The system (\ref{eq:eq_balance_equations_compact}) and (\ref{eq:eq_boundary_conditions_compact}) is a two-point boundary value problem. The problem can readily be solved by a Chebyschev collocation method using the MATLAB package \textit{Chebfun} \cite{driscoll2014chebfun}. 

\section{Simulations and Results}

In the following, we will investigate the influences of system parameters, especially length ratio $\lambda_l$, bending stiffness ratio $\lambda_B$, and line density ratio $\lambda_m$, upon the performance of the proposed PEHEE. These three parameters are determined by the length $l_e$, Young's modulus $Y_e$, and volumetric density $\rho_e$ of the extension beam respectively. Note that these parameters are actually changing across different sets of simulations in the following subsections. In the simulation, base excitation frequency $f_b$ and external load resistance $R_l$, which change the dimensionless values of $\nu$ and $\beta$, respectively, are of critical importance in the sense that these two parameters reflect the influence of vibration source and external load circuit. For every set of parameter values, we set the base excitation frequency $f_b$ to change from $1\ Hz$ to $100\ Hz$, which covers the usual frequency range of natural vibration sources, and set the load resistance $R_l$ to change from $1\ \Omega$ to $10\ M\Omega$, which is inspired by the Ref. \cite{erturk2009experimentally} and takes into account the dielectric property of piezoelectric materials. Actually, when the load resistance $R_l = 1\ \Omega$, the PEHEE is said to be in a short-circuit condition as the equivalent resistance of the PEHEE is generally much larger than $R_l$. On the other hand, when $R_l = 10\ M\Omega$, the system is close to an open-circuit condition where no external load is connected to the output electrodes.

\begin{table}[!htbp]
    \caption{Geometric and material parameters used in the simulation for PEHEEs}
    \label{tab:parameter_value_extension}
    \centering
    \begin{tabular}{lc}
    \hline
    \hline
    \textbf{Parameter} & \textbf{Value} \\
    \hline
       Length of the primary beam, $l_p$ $(mm)$  & 100 \\
       Width of the PEHEE, $b$ $(mm)$  & 20 \\
       Half thickness of the base structure, $h_s$ $(mm)$  & 0.25 \\
       Thickness of the piezoelectric element, $h_p$ $(mm)$  & 0.2 \\
       Young's modulus of the base structure, $Y_s$ $(Gpa)$  & 100 \\
       Young's modulus of the piezoelectric element, $Y_p$ $(Gpa)$  & 66 \\
       Mass density of the base structure, $\rho_s$ $(kg/m^3)$  & 7165 \\
       Mass density of the piezoelectric element, $\rho_p$ $(kg/m^3)$  & 7800 \\
       piezoelectric constant, $d_{31}$ $(pm/V)$  & -190 \\
       Permittivity, $\epsilon^S_{33}$ $(nF/m)$  & 15.93 \\
       Length of the extension beam, $l_e$ $(mm)$  & 100 \\
       Young's modulus of the extension beam, $Y_e$ $(Gpa)$  & 2.3 \\
       Mass density of the extension beam, $\rho_e$ $(kg/m^3)$  & 1.38 \\
       Half thickness of the extension beam, $h_e$ $(mm)$  & 0.25 \\
    \hline
    \hline
    \end{tabular}
\end{table}

\subsection{Base excitation response of the PEHEE}

\begin{figure}[!htbp]
    \centering
    \includegraphics[width=\textwidth]{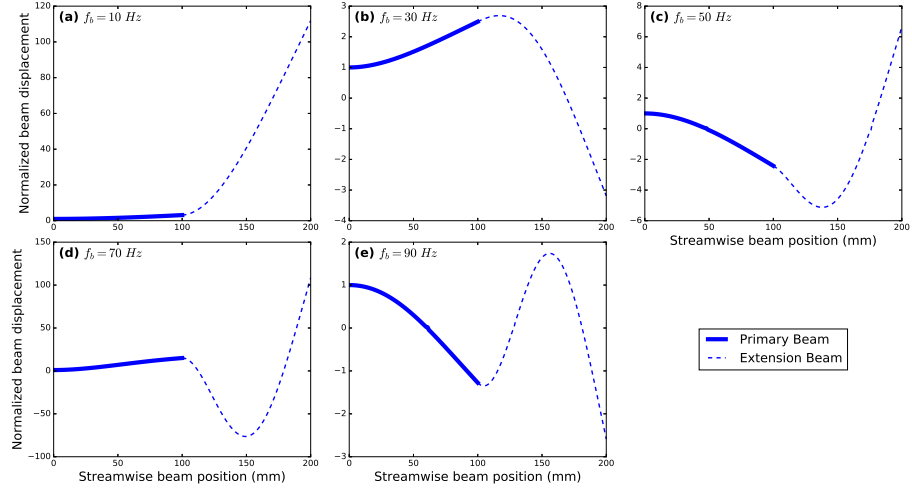}
    \caption{Profile of the PEHEE vibration at different base excitation frequency $f_b$. The beam displacement function is normalized with respect to the base excitation amplitude $\xi_b$. }
    \label{fig:fig_vibration_profile_vs_fr_Rl_laml_all}
\end{figure}

Firstly, we investigate the response of the PEHEE under base excitation. The basic geometry and material properties of the materials used are summarized in Table~\ref{tab:parameter_value_extension}. By using the MATLAB package \textit{Chebfun} \cite{driscoll2014chebfun}, we calculate the solution of the previously described boundary value problem with respect to different values of base excitation frequency $f_b$ and load resistance $R_l$. For the case of $R_l = 10\ k\Omega$, we plot the PEHEE vibration profile at different base excitation frequency $f_b$ in Figure~\ref{fig:fig_vibration_profile_vs_fr_Rl_laml_all}.

From the figure, it is clearly seen that if we consider the PEHEE as a whole, its vibration is extremely similar to that of a CPEH. \cite{erturk2009effect} With the increase of base excitation frequency $f_b$, more and more strain nodes are presented. Nonetheless, the PEHEE can also be regarded as a combination of the primary beam and the extension beam. In this view, we see from Figure~\ref{fig:fig_vibration_profile_vs_fr_Rl_laml_all} that in the considered frequency range, the vibration profile of the primary beam is always monotone without showing any strain nodes. For the extension beam, depending on the base excitation frequency $f_b$, one or more strain nodes are presented. Therefore, we can conclude that the overall motion of the PEHEE is a result of the interaction between the primary beam and the extension beam. 

\subsection{Influence of length $l_e$ of the extension beam or length ratio $\lambda_l$}

The presence of the extension beam is primarily indicated by the extension beam length $l_e$, or equivalently the length ratio $\lambda_l$. When $\lambda_l = 0.0$, no extension beam is attached to the primary beam and the resultant piezoelectric energy harvester reduces to a CPEH \cite{erturk2009experimentally}, which is referred to as a reference for comparison. In this contribution, the range of length ratio $\lambda_l$ to be considered is $0.0 \leq \lambda_l \leq 1.0$. For each value of length ratio $\lambda_l$, the above described base excitation problem is solved with respect to different base excitation frequency $f_b$ and load resistance $R_l$. In this process, Young's modulus of the extension beam is set to be $Y_e = 2.3\ GPa$ while its volumetric density is set to be $\rho_e = 1.38\times10^3\ kg/m^3$.

\begin{figure}[!htbp]
    \centering
    \includegraphics[width=\textwidth]{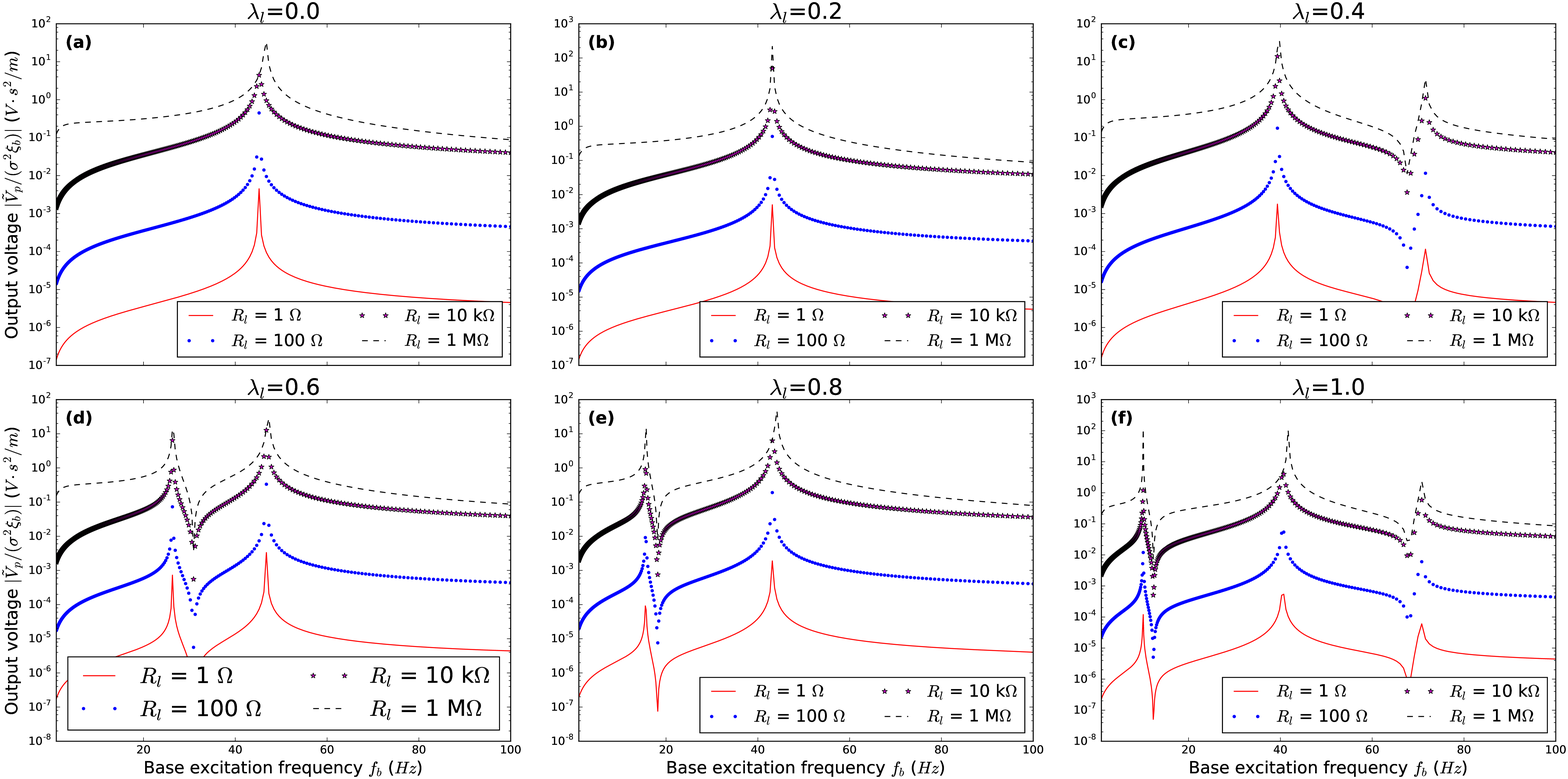}
    \caption{ Normalized output voltage of the PEHEE versus base excitation frequency $f_b$ at different length ratio $\lambda_l$ and load resistance $R_l$. }
    \label{fig:fig_output_voltage_vs_fr_Rl_laml_all}
\end{figure}

\begin{figure}[!htbp]
    \centering
    \includegraphics[width=\textwidth]{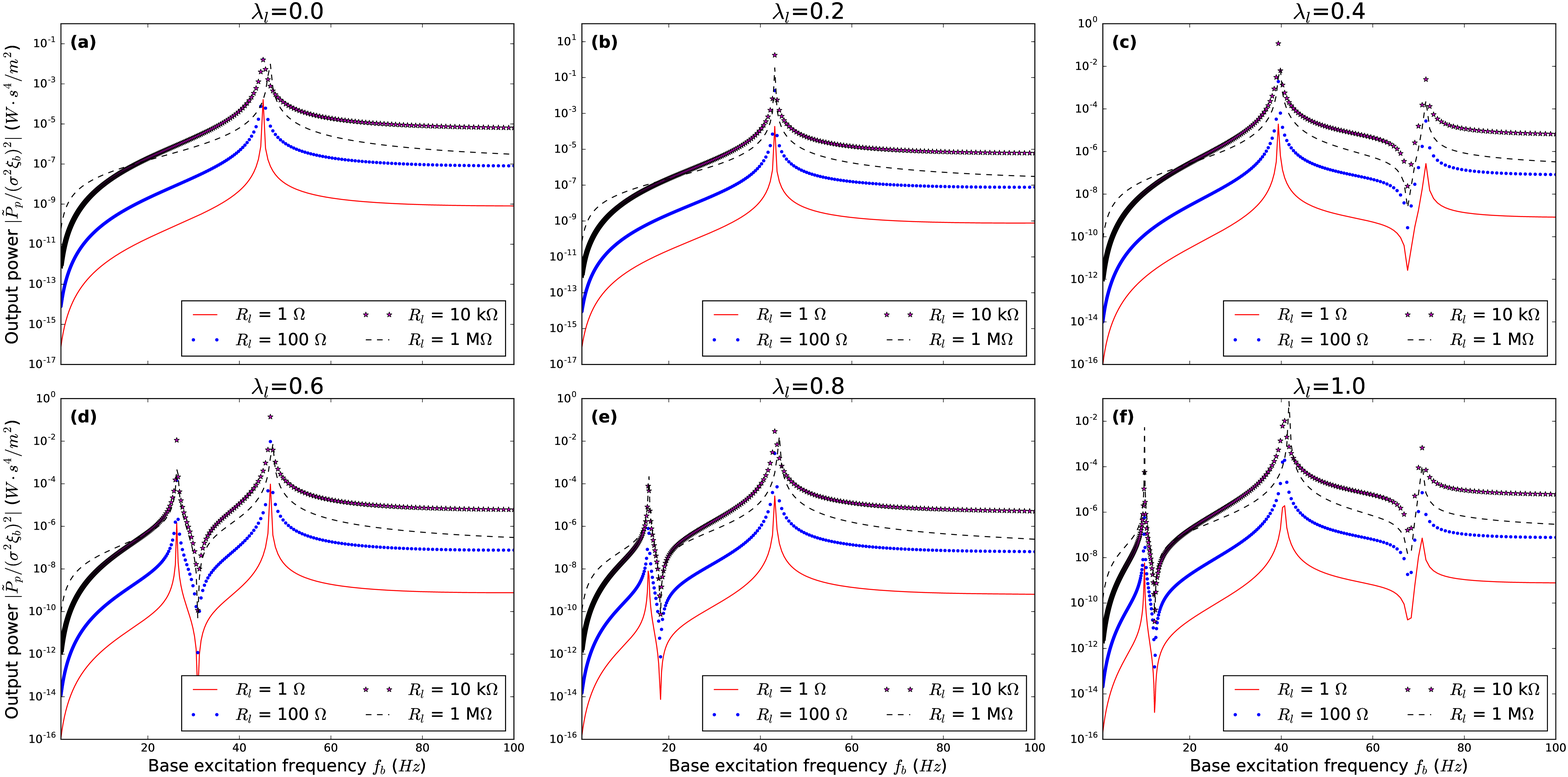}
    \caption{ Normalized output power of the PEHEE versus base excitation frequency $f_b$ at different length ratio $\lambda_l$ and load resistance $R_l$. }
    \label{fig:fig_output_power_vs_fr_Rl_laml_all}
\end{figure}

In the first place, we would like to explore the influence of base excitation frequency $f_b$ upon performance of the PEHEE at different values of load resistance $R_l$ and length ratio $\lambda_l$. Let the frequency range of interest be $1 - 100\ Hz$. The length ratio $\lambda_l$ is chosen to be $0.0$, $0.2$, $0.4$, $0.6$, $0.8$, or $1.0$, and load resistance $R_l$ is chosen to be $10^0\ \Omega$, $10^2\ \Omega$, $10^4\ \Omega$, or $10^6\ \Omega$. According the the previous sections, output voltage $\tilde{V}_p$ and output power $\tilde{P}_p$ can be evaluated by
\begin{equation}
    \begin{aligned}
        \tilde{V}_p &= \frac{j \nu \beta}{j \nu \beta + 1} \frac{\xi_b}{l_p} \frac{e_p}{C_p}, \\
        \tilde{P}_p &=  \tilde{V}_p^2 / R_l.
    \end{aligned}
\end{equation}
Note that here $\tilde{V}_p$ and $\tilde{P}_p$ are the complex amplitudes of the periodic output voltage $V_p(t)$ and output power $P_p(t)$, which contain the modulus and phase information. Another point to be noted is that the frequencies of $V_p(t)$ and $P_p(t)$ are actually different. The frequency of $P_p(t)$ is twice as much as that of $V_p(t)$. In this contribution, we are more interested in the modulus information, which are $|\tilde{V}_p|$ and $|\tilde{P}_p|$, respectively. The simulation results of $|\tilde{V}_p|$ and $|\tilde{P}_p|$ are plotted against base excitation frequency $f_b$ in Figure~\ref{fig:fig_output_voltage_vs_fr_Rl_laml_all} and Figure~\ref{fig:fig_output_power_vs_fr_Rl_laml_all}, respectively. In the plots the output voltage is normalized with respect to the amplitude of base excitation amplitude $\xi_b \sigma^2$, while the output power is normalized with respect to $\left(\xi_b \sigma^2\right)^2 = \xi_b^2 \sigma^4$. In this process, the values of density $\rho_e$ and Young's modulus $Y_e$ of the extension beam, or equivalently $\lambda_m$ and $\lambda_B$, are kept unchanged.

\begin{figure}[!htbp]
    \centering
    \includegraphics[width=\textwidth]{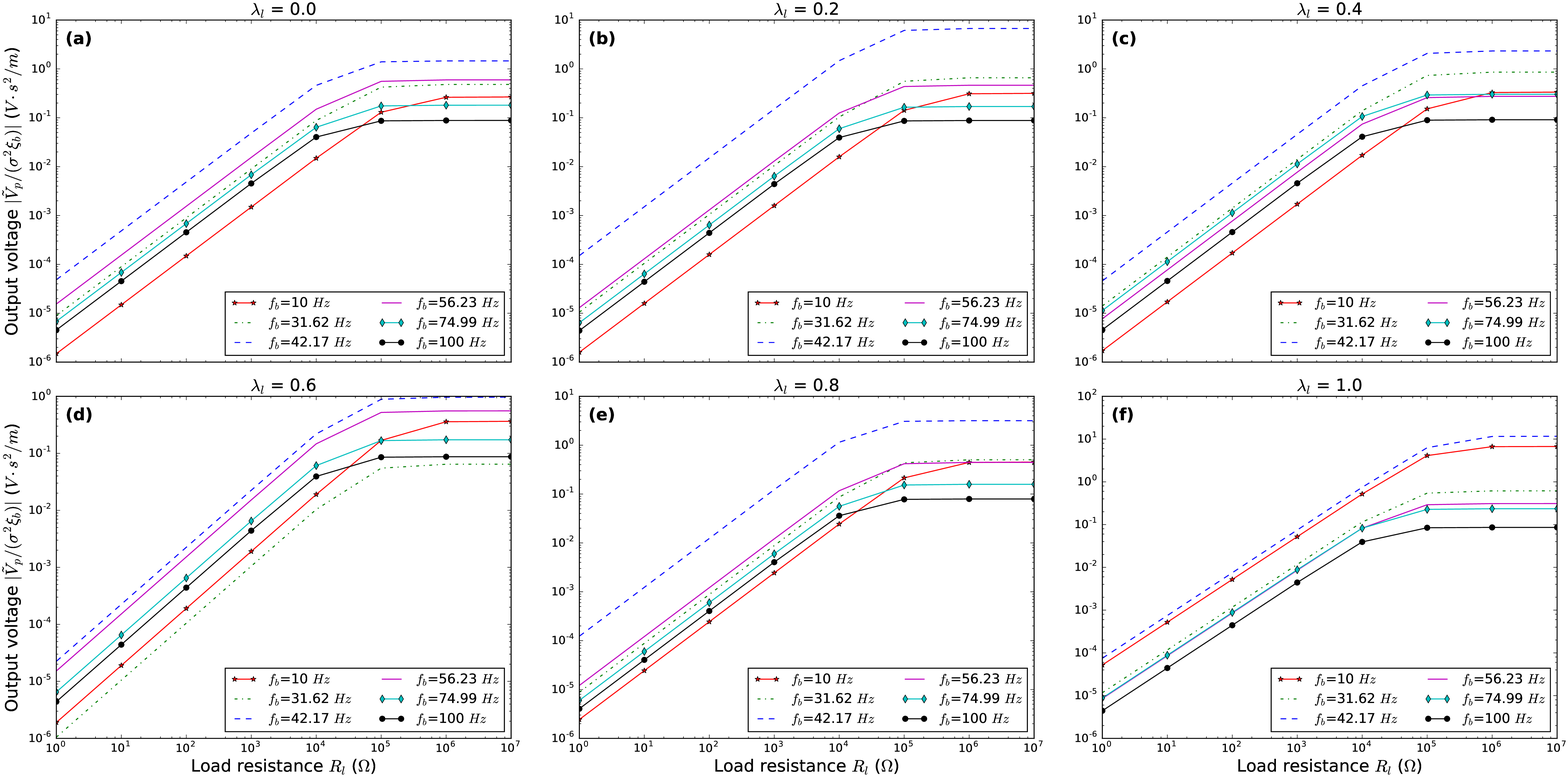}
    \caption{ Normalized output voltage of the PEHEE versus load resistance $R_l$ at different base excitation frequency $f_b$ and length ratio $\lambda_l$. }
    \label{fig:fig_vol_laml_list_vs_fr_Rl}
\end{figure}

\begin{figure}[!htbp]
    \centering
    \includegraphics[width=\textwidth]{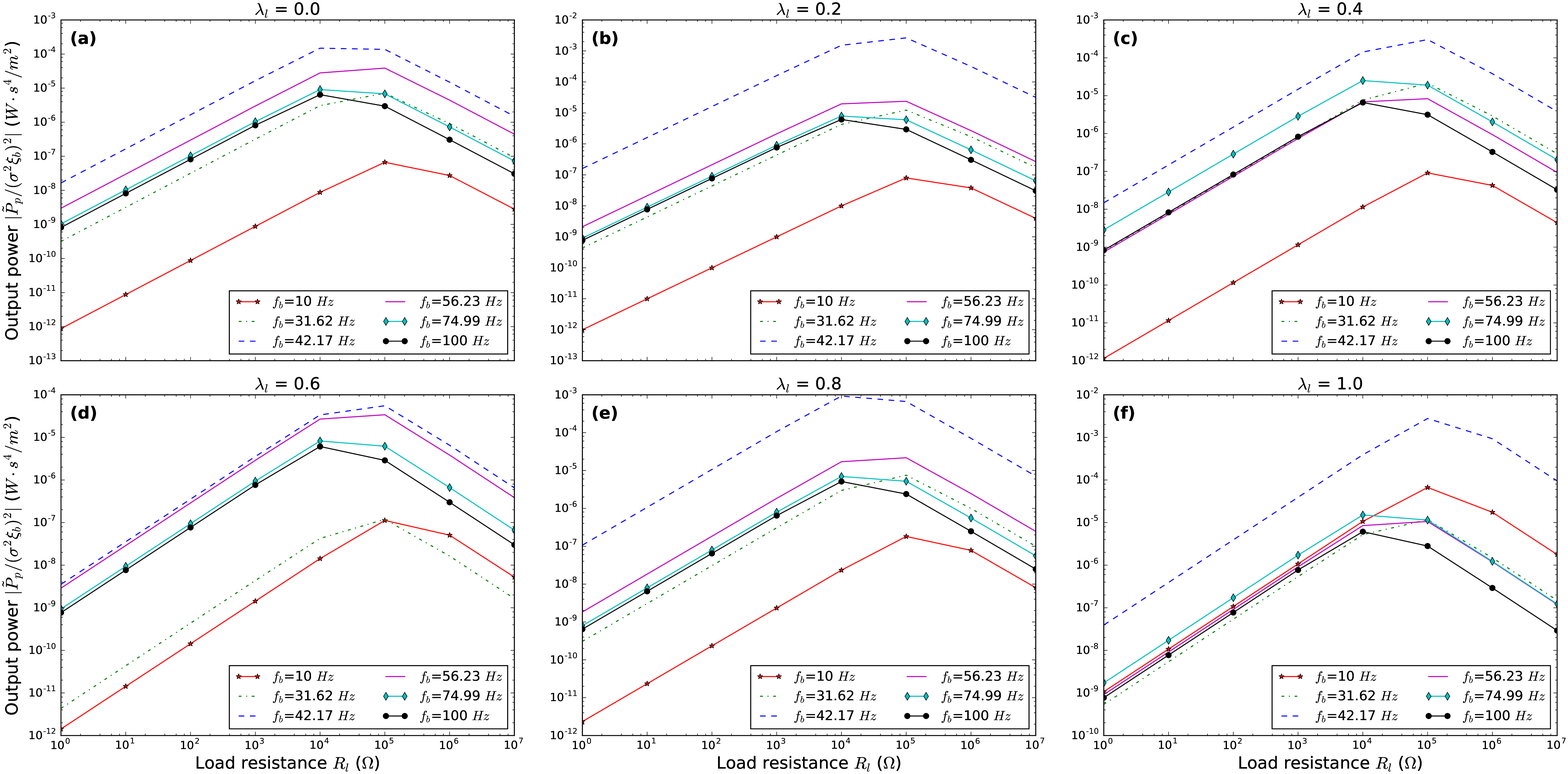}
    \caption{ Normalized output power of the PEHEE versus load resistance $R_l$ at different base excitation frequency $f_b$ and length ratio $\lambda_l$. }
    \label{fig:fig_pow_laml_list_vs_fr_Rl}
\end{figure}

According to Figure~\ref{fig:fig_output_voltage_vs_fr_Rl_laml_all} and Figure~\ref{fig:fig_output_power_vs_fr_Rl_laml_all}, for the given values $\lambda_m$ and $\lambda_B$, frequency response of the PEHEE in the interested frequency range changes in accordance with the extension length $l_e$ and thus the length ratio $\lambda_l$. When $\lambda_l$ is relatively small, say $\lambda_l = 0.2$, as shown in Figure~\ref{fig:fig_output_voltage_vs_fr_Rl_laml_all}(b), frequency response of the PEHEE is almost the same as that of the CPEH case where $\lambda_l = 0.0$. That is to say, there is only one resonant peak in the considered frequency range and the corresponding resonant frequency is around $42\ Hz$ for different values of load resistance $R_l$. With the increase of the length ratio $\lambda_l$, say $\lambda_l = 0.4$, $0.6$, or $0.8$, as shown in Figure~\ref{fig:fig_output_voltage_vs_fr_Rl_laml_all}(c), (d), and (e), respectively, there begins to exist an extra resonant and anti-resonant mode in the considered frequency range, compared with the case of no elastic extension ($\lambda_l=0.0$). If we further increase the length ratio to $\lambda_l = 1.0$, more resonant and anti-resonant modes occur in the considered frequency range. It can be found from Figure~\ref{fig:fig_output_voltage_vs_fr_Rl_laml_all}(d) and Figure~\ref{fig:fig_output_power_vs_fr_Rl_laml_all}(d) that, amplitude of the output voltage $\tilde{V}_p$ and output power $\tilde{P}_p$ at the newly occurring resonant mode is comparable in order of magnitude to that of the CPEH. 

Since the primary beam is much stiffer than the extension beam in our case, the above results can be explained as follows. As indicated before, the overall motion of the PEHEE is governed by the interaction between the primary beam and extension beam. When the length ratio $\lambda_l$ is small, the interaction is too small to play a role in the motion of the whole PEHEE. Thus, the motion of the whole PEHEE is mainly determined by that of the primary beam and its frequency response is therefore similar to that of a CPEH. When the length ratio $\lambda_l$ is increased to a level such that the interaction between the primary beam and the extension is no longer negligible, the motion of the whole PEHEE is starting to be influenced by the extension beam. A direct result is that the resonant frequency of the resonant peak is shifted by a small amount, as shown in Figure~\ref{fig:fig_output_voltage_vs_fr_Rl_laml_all}(c) and (d), where $\lambda_l = 0.4$ and $0.6$ respectively. A second result is that the newly occurring resonant and anti-resonant mode possesses a resonant frequency that decreases with respect to the increase of $\lambda_l$. After a critical value, the further increase of $\lambda_l$ leads to the weakening of the interaction again, as shown in Figure~\ref{fig:fig_output_voltage_vs_fr_Rl_laml_all}(e) where $\lambda = 0.8$. In this case, the two parts again have little effect upon each other the performance of the PEHEE is again governed by the primary beam. It should be noted that there can be more resonant and anti-resonant modes introduced into the considered frequency range if we further increase the value of $\lambda_l$, as shown in Figure~\ref{fig:fig_output_voltage_vs_fr_Rl_laml_all}(f). The newly introduced mode is mainly determined by the extension beam and shows energy harvesting potential as indicated above. 

\begin{figure}[!htbp]
    \centering
    \includegraphics[width=\textwidth]{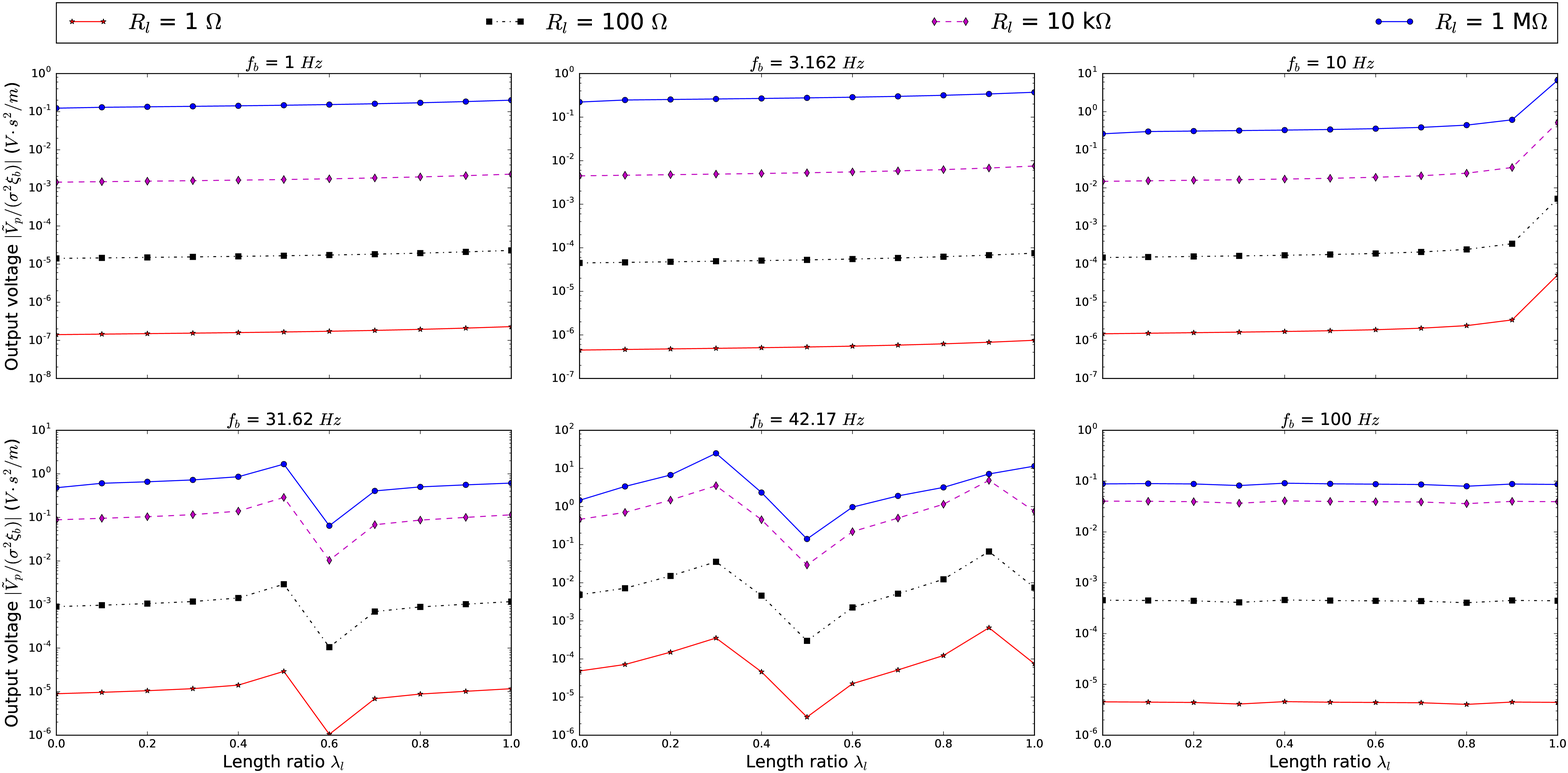}
    \caption{ Normalized output voltage of the PEHEE versus length ratio $\lambda_l$ at different base excitation frequency $f_b$ and load resistance $R_l$. }
    \label{fig:fig_vol_fr_sl_Rl_sl_vs_laml}
\end{figure}

\begin{figure}[!htbp]
    \centering
    \includegraphics[width=\textwidth]{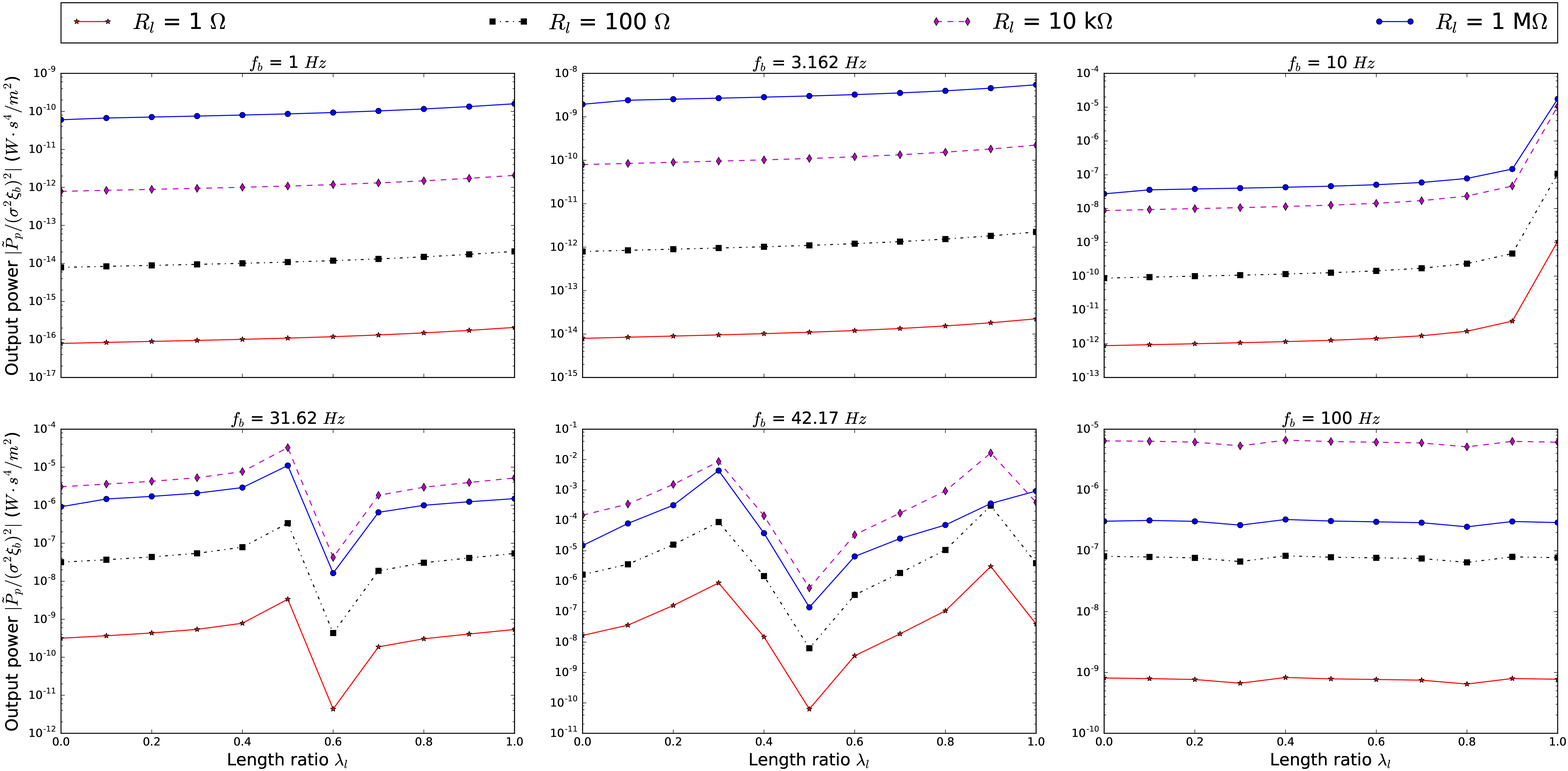}
    \caption{ Normalized output power of the PEHEE versus length ratio $\lambda_l$ at different base excitation frequency $f_b$ and load resistance $R_l$. }
    \label{fig:fig_pow_fr_sl_Rl_sl_vs_laml}
\end{figure}

In this view, attachment of the elastic extension part actually increases the working frequency range of the proposed PEHEE, as typical piezoelectric energy harvesters rely on resonant modes to work. On the other hand, the elastic extension also provides a way to tune the bandwidth of a piezoelectric energy harvester. An obvious way to do so is to choose the parameters of the extension beam, so that the extension beam and the primary beam have similar resonant frequency and interact strongly with each other. The point to be noticed is the existence of anti-resonant modes. They correspond to low output voltage and output power, and to some degree narrows the bandwidth of the energy harvester. However, it is still not worse than an otherwise non-resonant vibration, as in both cases the electrical output of the piezoelectric energy harvester is unusable.

In the second place, we investigate the influence of load resistance $R_l$ upon performance of the PEHEE at different values of base excitation frequency $f_b$ and length ratio $\lambda_l$. The resulting normalized output voltage and output power are shown in Figure~\ref{fig:fig_vol_laml_list_vs_fr_Rl} and Figure~\ref{fig:fig_pow_laml_list_vs_fr_Rl}, respectively. It is obvious that in the range of low $R_l$, a power law exists between the normalized output voltage $|\tilde{V}_p/(\sigma^2 \xi_b)|$ and normalized output power $|\tilde{P}_p/(\sigma^2 \xi_b)^2|$ and load resistance $R_l$. For all the chosen values of base excitation frequency $f_b$, the increase of $R_l$ leads to an increase in $|\tilde{V}_p/(\sigma^2 \xi_b)|$. Ultimately, $|\tilde{V}_p/(\sigma^2 \xi_b)|$ approaches asymptotically to a limit value $\tilde{V}_p^{lim}$. This value actually corresponds to the amplitude of open-circuit output voltage of the piezoelectric energy harvester. At the same time, the value of $|\tilde{P}_p/(\sigma^2 \xi_b)^2|$ exhibits an obvious maximum at some values of $R_l$ between $10\ k\Omega$ and $1\ M\Omega$. And on both sides away from the maximum value, we see also a power law between the value of $R_l$ and $|\tilde{P}_p/(\sigma^2 \xi_b)^2|$, as shown in Figure~\ref{fig:fig_pow_laml_list_vs_fr_Rl}. This indicates that an asymptotic analysis may help us to simplify the analysis of performance of piezoelectric energy harvesters and provide a more profound understanding of piezoelectric energy harvesters. But this is out of the scope of our current contribution and will be covered in the future research.

In the third place, at different values of $f_b$ and $R_l$, we directly plot the normalized output voltage $|\tilde{V}_p/(\sigma^2 \xi_b)|$ and normalized output power $|\tilde{P}_p/(\sigma^2 \xi_b)^2|$ with respect to length ratio $\lambda_l$, which are shown in Figure~\ref{fig:fig_vol_fr_sl_Rl_sl_vs_laml} and Figure~\ref{fig:fig_vol_fr_sl_Rl_sl_vs_laml}, respectively. It is shown that at given values of $R_l$ and $f_b$, output performances of the proposed PEHEE can be tuned by the parameter $\lambda_l$. For example, when $f_b = 31.62\ Hz$ and $R_l = 10\ k\Omega$, a maximum peak of $|\tilde{V}_p/(\sigma^2 \xi_b)|$ and $|\tilde{P}_p/(\sigma^2 \xi_b)^2|$ is found around $\lambda_l = 0.5$. Compared with the case of no elastic extension ($\lambda_l = 0.0$), the amplitude of output voltage $|\tilde{V}_p/(\sigma^2 \xi_b)|$ is about $13$ times larger, while the amplitude of output power $|\tilde{P}_p/(\sigma^2 \xi_b)^2|$ is about $3$ times larger. Similar phenomena can be found when $f_b = 42.17\ Hz$. This indicates that the addition of the elastic extension can substantially increase the output performance of a piezoelectric energy harvester. It should be noted that in  case of large values of $f_b$, like $f_b = 100\ Hz$, the tuning performance of the elastic extension is no longer significant.

\subsection{Influence of Young's modulus $Y_e$ of the extension beam or bending stiffness ratio $\lambda_B$}

To investigate the influence of bending stiffness ratio $\lambda_B$ upon the performance of the PEHEE, we set different values of Young's modulus $Y_e$ of the extension beam to change the values of $\lambda_B$. Considering the properties of commonly used engineering materials \cite{warlimont2018springer}, the values of $Y_e$ is set to be in the range of $0.01\ GPa \leq Y_e \leq 400\ GPa$. For each value of bending stiffness ratio $\lambda_B$, the base excitation problem is solved with respect to different base excitation frequency $f_b$ and load resistance $R_l$. In this process, the length ratio is fixed to be $\lambda_l = 0.3$ while its volumetric density is set to be $\rho_e = 1.38 \times 10^3\ kg/m^3$.

\begin{figure}[!htbp]
    \centering
    \includegraphics[width=\textwidth]{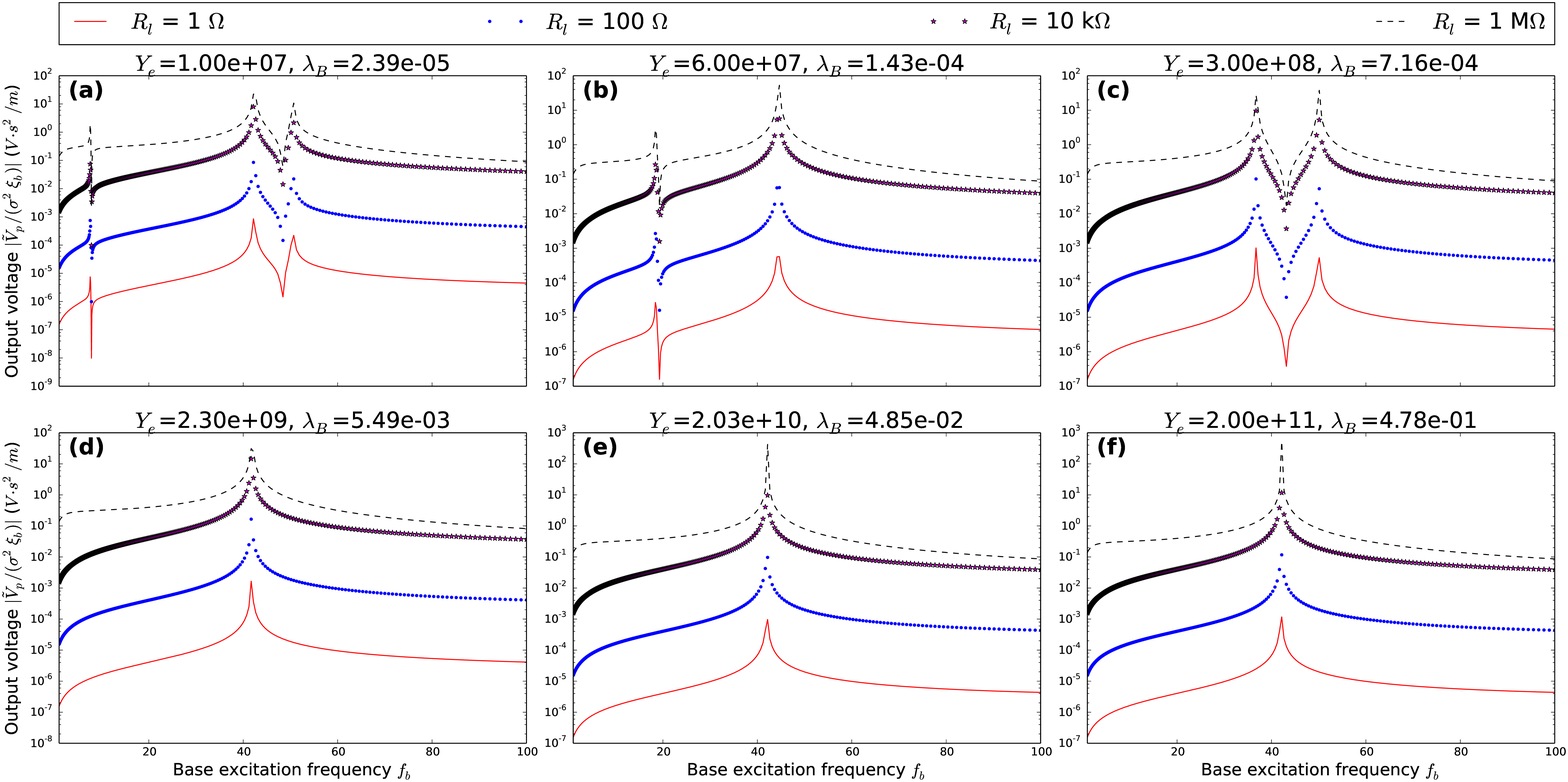}
    \caption{ Normalized output voltage of the PEHEE versus base excitation frequency $f_b$ at different bending stiffness ratio $\lambda_B$ and load resistance $R_l$. }
    \label{fig:fig_output_voltage_vs_fr_Rl_lamB_all}
\end{figure}

\begin{figure}[!htbp]
    \centering
    \includegraphics[width=\textwidth]{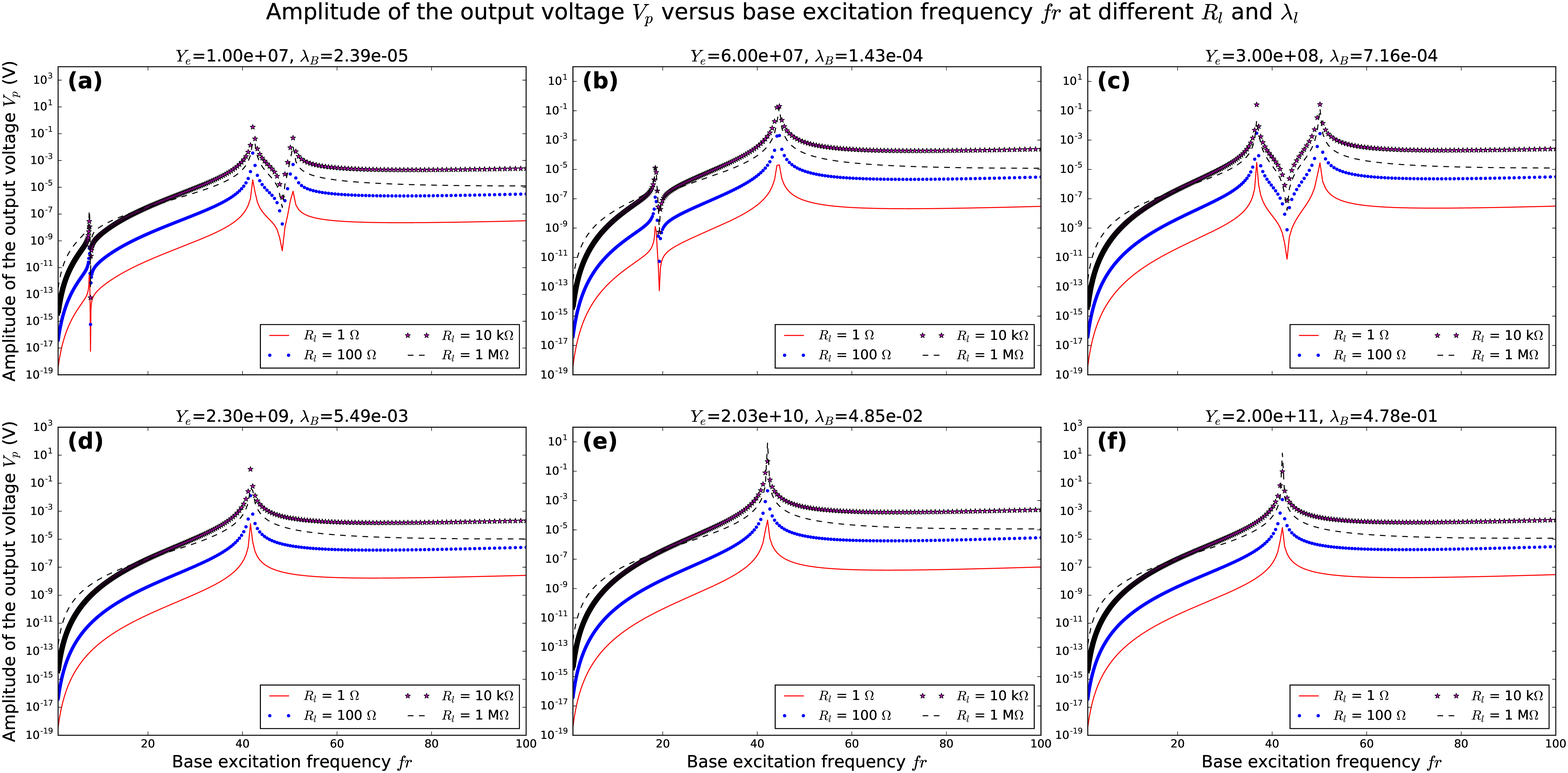}
    \caption{Normalized output power of the PEHEE versus base excitation frequency $f_b$ at different bending stiffness ratio $\lambda_B$ and load resistance $R_l$.  }
    \label{fig:fig_output_power_vs_fr_Rl_lamB_all}
\end{figure}

Firstly, the influence of base excitation frequency $f_b$ upon performance of the piezoelectric energy harvester is shown Figure~\ref{fig:fig_output_voltage_vs_fr_Rl_lamB_all} and Figure~\ref{fig:fig_output_power_vs_fr_Rl_lamB_all} at different values of load resistance $R_l$ and bending stiffness ratio $\lambda_B$. To have a clear clue about the value of $\lambda_B$, it is found that for a Young's modulus of $Y_e = 0.01\ GPa$, the corresponding value of bending stiffness ratio is $\lambda_B = 2.39\times10^{-5}$, while for a Young's modulus of $Y_e = 200\ GPa$, the corresponding value of bending stiffness ratio is $\lambda_B = 0.478$.

It is easily seen from Figure~\ref{fig:fig_output_voltage_vs_fr_Rl_lamB_all} and Figure~\ref{fig:fig_output_power_vs_fr_Rl_lamB_all} that, at the given values of $\lambda_l$ and $\lambda_m$, bending stiffness ratio $\lambda_B$ also affects frequency response of the proposed PEHEE, but in a different manner from length ratio $\lambda_l$.

\begin{figure}[!htbp]
    \centering
    \includegraphics[width=\textwidth]{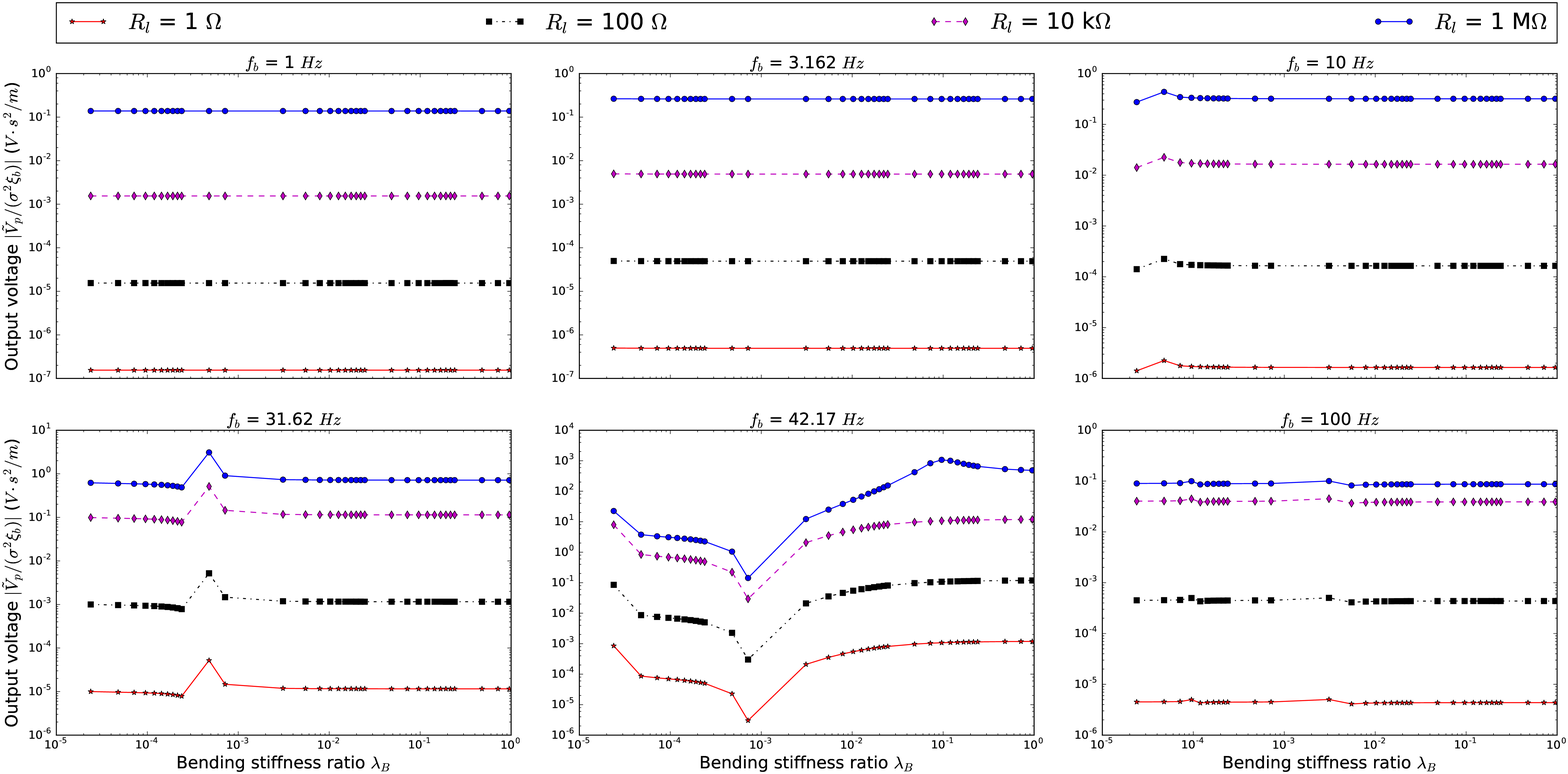}
    \caption{ Normalized output voltage of the PEHEE versus bending stiffness ratio $\lambda_B$ at different base excitation frequency $f_b$ and load resistance $R_l$. }
    \label{fig:fig_vol_fr_sl_Rl_sl_vs_lamB}
\end{figure}

\begin{figure}[!htbp]
    \centering
    \includegraphics[width=\textwidth]{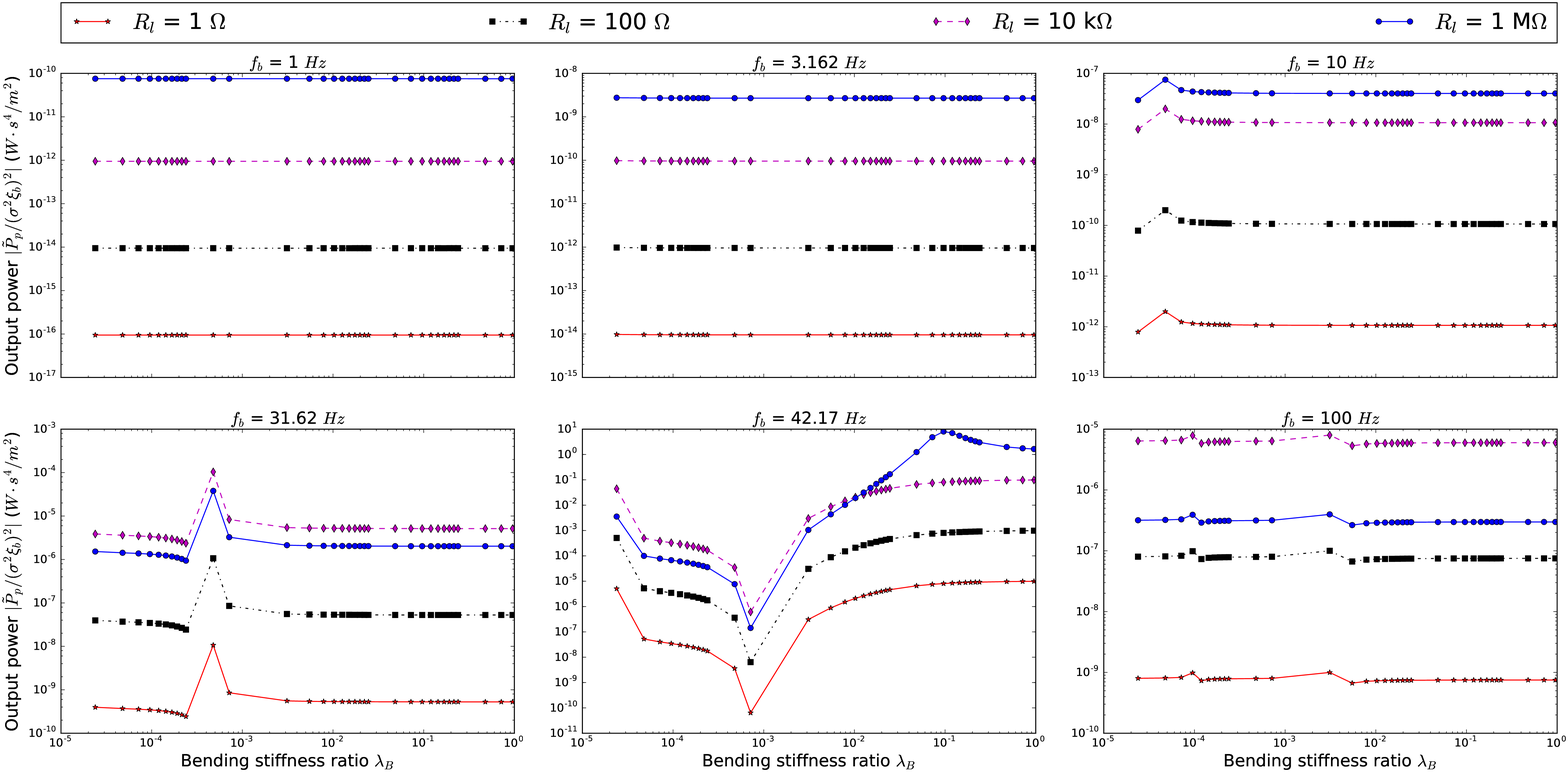}
    \caption{ Normalized output power of the PEHEE versus bending stiffness ratio $\lambda_B$ at different base excitation frequency $f_b$ and load resistance $R_l$. }
    \label{fig:fig_pow_fr_sl_Rl_sl_vs_lamB}
\end{figure}

When the bending stiffness ratio $\lambda_B$ is very small, say $Y_e = 0.01\ GPa$, as shown in Figure~\ref{fig:fig_output_voltage_vs_fr_Rl_lamB_all}(a) and Figure~\ref{fig:fig_output_power_vs_fr_Rl_lamB_all}(a), there exist multiple resonant modes in the considered frequency range. And some of the resonant modes are potential to be used for energy harvesting. With the increase of bending stiffness ratio $\lambda_B$, fewer resonant modes are present in the considered frequency range. For example, when $Y_e = 0.3\ GPa$, two resonant modes are present and when $Y_e = 2.3\ GPa$, only one resonant mode is present. As a result, less frequency points can be utilized for energy harvesting. That is to say, to achieve an energy harvesting capacity of wider frequency range, smaller bending stiffness ratio $\lambda_B$ is preferred. Actually, from Figure~\ref{fig:fig_output_voltage_vs_fr_Rl_lamB_all} we can conclude that, a bending stiffness ratio $\lambda_B$ in the order of $10^{-4}$ or lower is beneficial to the energy harvesting performance of the PEHEE. However, only changing the value of $\lambda_B$ is not the best way to tune the energy harvesting performance of the PEHEE. It is seen from the previous subsection and this subsection that, the combination change of larger $\lambda_l$ and lower $\lambda_B$ serve better for this goal.

The influence of bending stiffness ratio $\lambda_B$ is further shown by plotting the normalized output voltage $|\tilde{V}_p/(\sigma^2 \xi_b)|$ and normalized output power $|\tilde{P}_p/(\sigma^2 \xi_b)^2|$ relative to $\lambda_B$ at different values of load resistance $R_l$ and base excitation frequency $f_b$, as shown in Figure~\ref{fig:fig_vol_fr_sl_Rl_sl_vs_lamB} and Figure~\ref{fig:fig_pow_fr_sl_Rl_sl_vs_lamB}, respectively. It is found that for a low frequency base excitation, as shown in Figure~\ref{fig:fig_vol_fr_sl_Rl_sl_vs_lamB}(a) and (b), no big difference is seen when we change the values of $\lambda_B$. The normalized output voltage $|\tilde{V}_p/(\sigma^2 \xi_b)|$ and normalized output power $|\tilde{P}_p/(\sigma^2 \xi_b)^2|$ remain at a low level and the proposed PEHEE can not be used for energy harvesting. For a higher base excitation frequency $f_b$, as shown in Figure~\ref{fig:fig_vol_fr_sl_Rl_sl_vs_lamB}(c) and (d) and Figure~\ref{fig:fig_pow_fr_sl_Rl_sl_vs_lamB}(c) and (d), a peak of $|\tilde{V}_p/(\sigma^2 \xi_b)|$ and $|\tilde{P}_p/(\sigma^2 \xi_b)^2|$ is found for certain value of $\lambda_B$. The output performance of the PEHEE is increased by changing the value of $\lambda_B$. For a further higher value of $f_b$, as shown in Figure~\ref{fig:fig_vol_fr_sl_Rl_sl_vs_lamB}(e), the value of $\lambda_B$ significantly changes $|\tilde{V}_p/(\sigma^2 \xi_b)|$ and $|\tilde{P}_p/(\sigma^2 \xi_b)^2|$. To achieve higher energy harvesting performance, higher (in the order of $10^{-1}$) or lower (in the order of $10^{-5}$) values of $\lambda_B$ are preferred against a moderate (in the order of $10^{-3}$) value of $\lambda_B$. For the value of base excitation frequency $f_b = 100\ Hz$, as shown in Figure~\ref{fig:fig_vol_fr_sl_Rl_sl_vs_lamB}(f), the value of $\lambda_B$ is found again playing a minor role in changing the values of $|\tilde{V}_p/(\sigma^2 \xi_b)|$ and $|\tilde{P}_p/(\sigma^2 \xi_b)^2|$.

\subsection{Influence of the density $\rho_e$ of the extension beam or line density ratio $\lambda_m$}

To investigate the influence of line density ratio $\lambda_m$ upon the performance of the PEHEE, the volumetric density $\rho_e$ of the beam extension part is changed based on the properties of commonly used engineering materials \cite{warlimont2018springer}. For the convenience of performance comparison, $\rho_e$ is set to be $1.38\times10^1\ kg/m^3$, $1.38\times10^2\ kg/m^3$, $1.38\times10^4\ kg/m^3$, or $1.38\times10^5\ kg/m^3$. For each value of line density ratio $\lambda_m$, the base excitation problem is solved for different base excitation frequency $f_b$ and load resistance $R_l$. During the simulation, the length ratio is fixed to be $\lambda_l = 0.3$ while the Young's modulus of the extension beam is set to be $Y_e = 2.3\ GPa$.

\begin{figure}[!htbp]
    \centering
    \includegraphics[width=\textwidth]{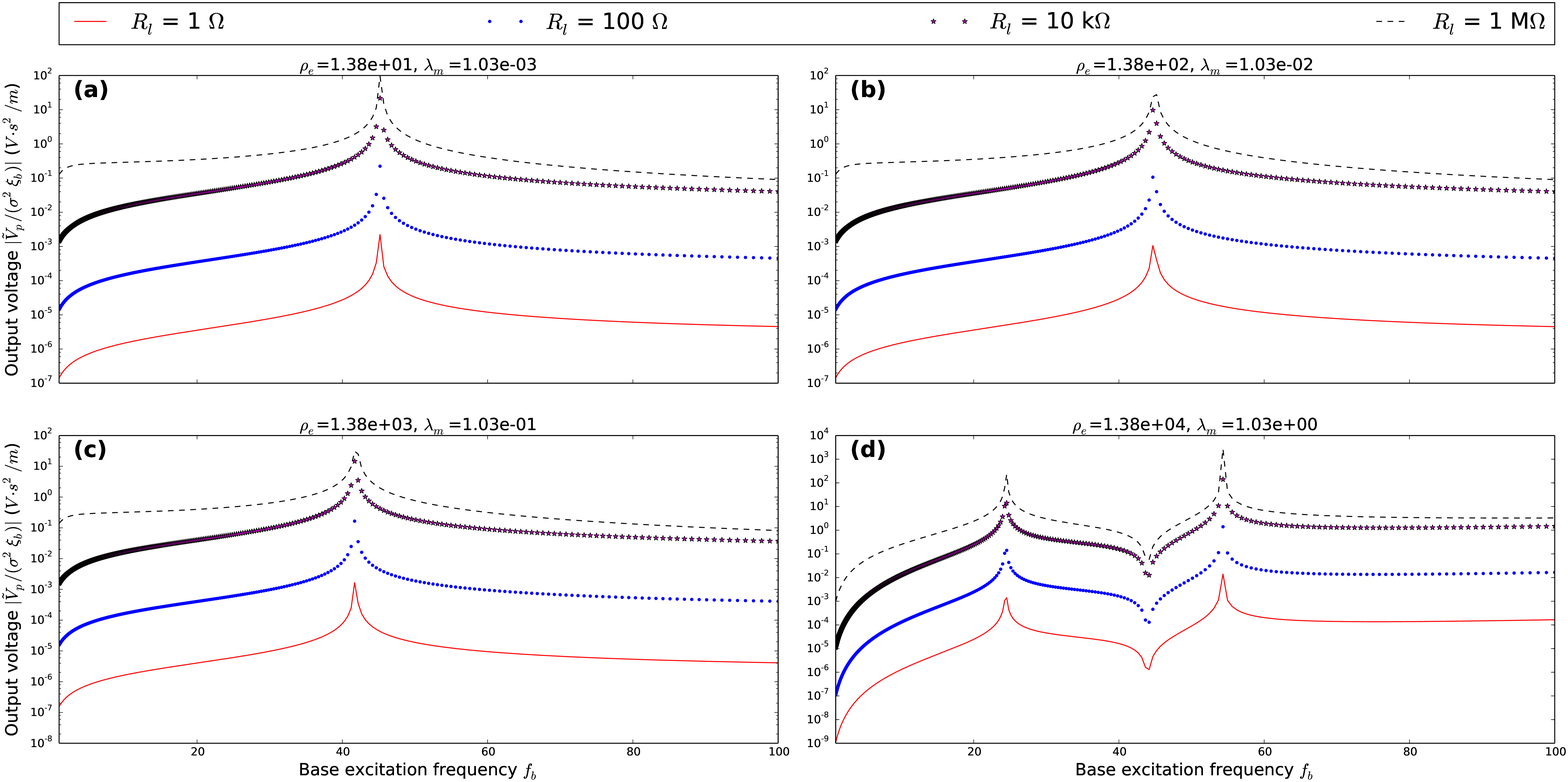}
    \caption{ Normalized output voltage of the PEHEE versus base excitation frequency $f_b$ at different line density ratio $\lambda_m$ and load resistance $R_l$.}
    \label{fig:fig_output_voltage_vs_fr_Rl_lamm_all}
\end{figure}

\begin{figure}[!htbp]
    \centering
    \includegraphics[width=\textwidth]{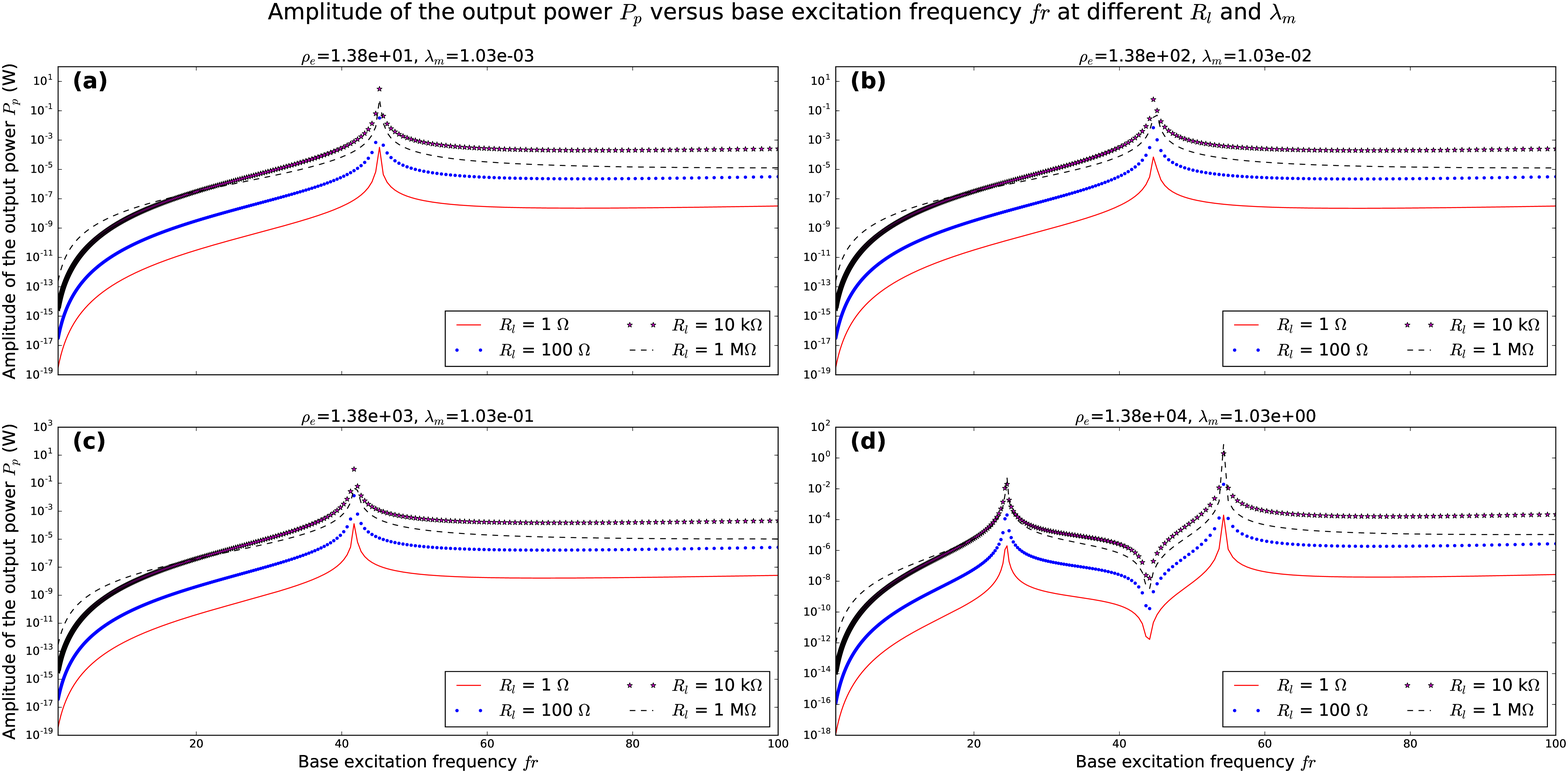}
    \caption{ Normalized output power of the PEHEE versus base excitation frequency $f_b$ at different line density ratio $\lambda_m$ and load resistance $R_l$.}
    \label{fig:fig_output_power_vs_fr_Rl_lamm_all}
\end{figure}

The influence of base excitation frequency $f_b$ upon performance of the proposed PEHEE is shown Figure~\ref{fig:fig_output_voltage_vs_fr_Rl_lamm_all} and Figure~\ref{fig:fig_output_power_vs_fr_Rl_lamm_all} at different values of load resistance $R_l$ and bending stiffness ratio $\lambda_B$. A simple calculation shows that when $\rho_e = 1.38\times10^1\ kg/m^3$, $\lambda_e = 1.03\times10^{-3}$, while for $\rho_e = 1.38\times10^4\ kg/m^3$, $\lambda_m = 1.03$. For typical flexible materials like $PET$\cite{dean1999lange}, the volumetric density $\rho_e$ is $1.38\times10^3\ kg/m^3$, the calculation results are shown in Figure~\ref{fig:fig_output_voltage_vs_fr_Rl_lamm_all}(c) and Figure~\ref{fig:fig_output_power_vs_fr_Rl_lamm_all}(c). It is indicated that no extra resonant modes are present in the considered frequency range. Besides, no big difference is observed in terms of the normalized output voltage $|\tilde{V}_p/(\sigma^2 \xi_b)|$ and normalized output power $|\tilde{P}_p/(\sigma^2 \xi_b)^2|$, compared with the cases of lower $\rho_e$, as shown in Figure~\ref{fig:fig_output_voltage_vs_fr_Rl_lamm_all}(a) and (b) and Figure~\ref{fig:fig_output_power_vs_fr_Rl_lamm_all}(a) and (b). However, for a larger value of volumetric density $\rho_e$, as shown in Figure~\ref{fig:fig_output_voltage_vs_fr_Rl_lamm_all}(d) and Figure~\ref{fig:fig_output_power_vs_fr_Rl_lamm_all}(d), the existence of extra resonant modes is seen. 

As a result, we can conclude that a heavier extension beam is beneficial for energy harvesting. In fact, in the limiting case of an elastic extension whose elastic constants is infinite, the extension beam is equivalent to a rigid attached tip mass. \cite{erturk2009experimentally} However, it is not always possible to find appropriate materials at a limited expense. For most structure materials available in engineering applications, the volumetric densities are not enough to generate extra resonant modes in the range. Hence, it is not wise to depend on the parameter $\rho_e$ merely to tune the performance of the PEHEE. 

\begin{figure}[!htbp]
    \centering
    \includegraphics[width=\textwidth]{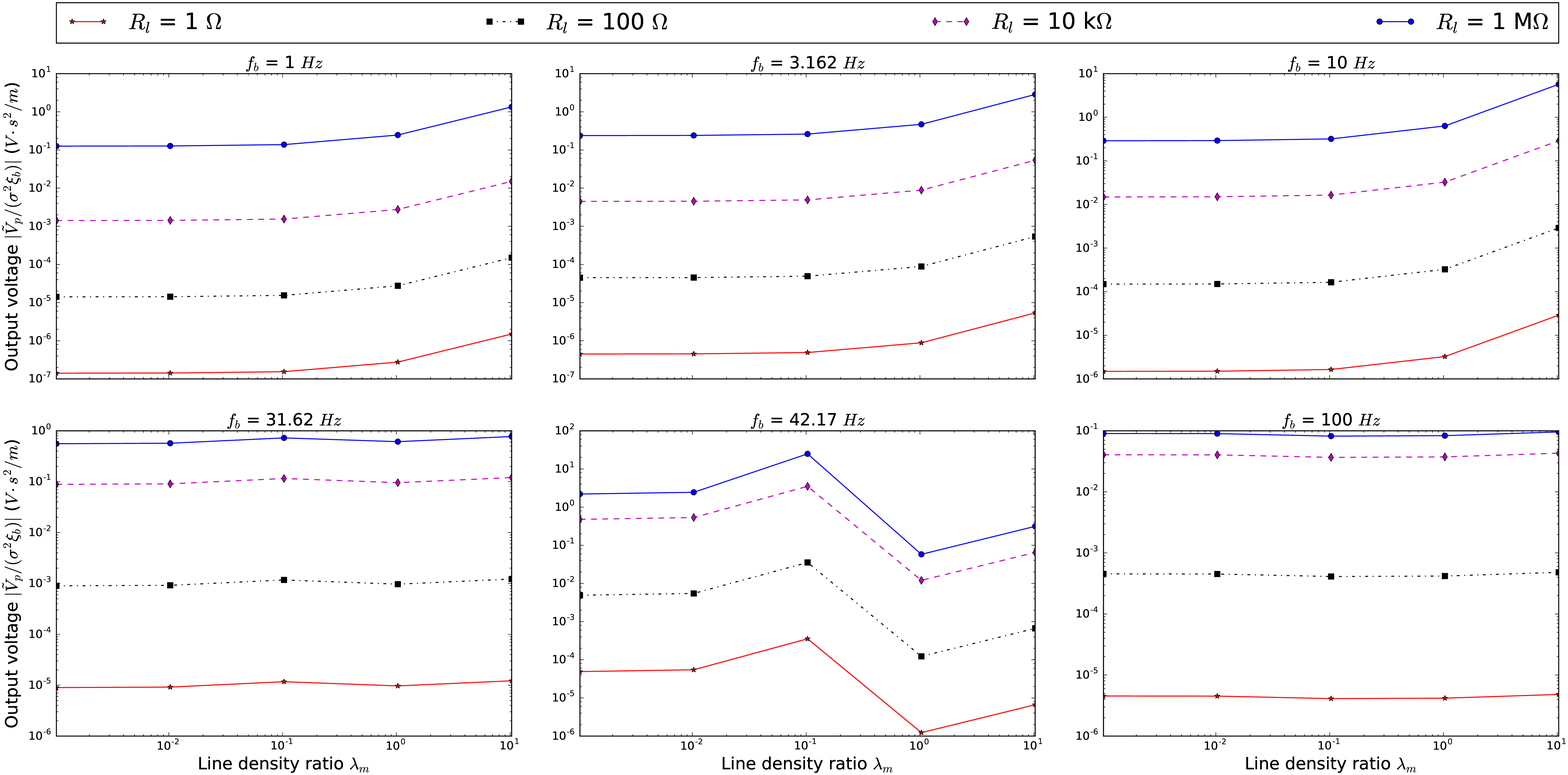}
    \caption{ Normalized output voltage of the PEHEE versus line density ratio $\lambda_m$ at different base excitation frequency $f_b$ and load resistance $R_l$. }
    \label{fig:fig_vol_fr_sl_Rl_sl_vs_lamm}
\end{figure}

\begin{figure}[!htbp]
    \centering
    \includegraphics[width=\textwidth]{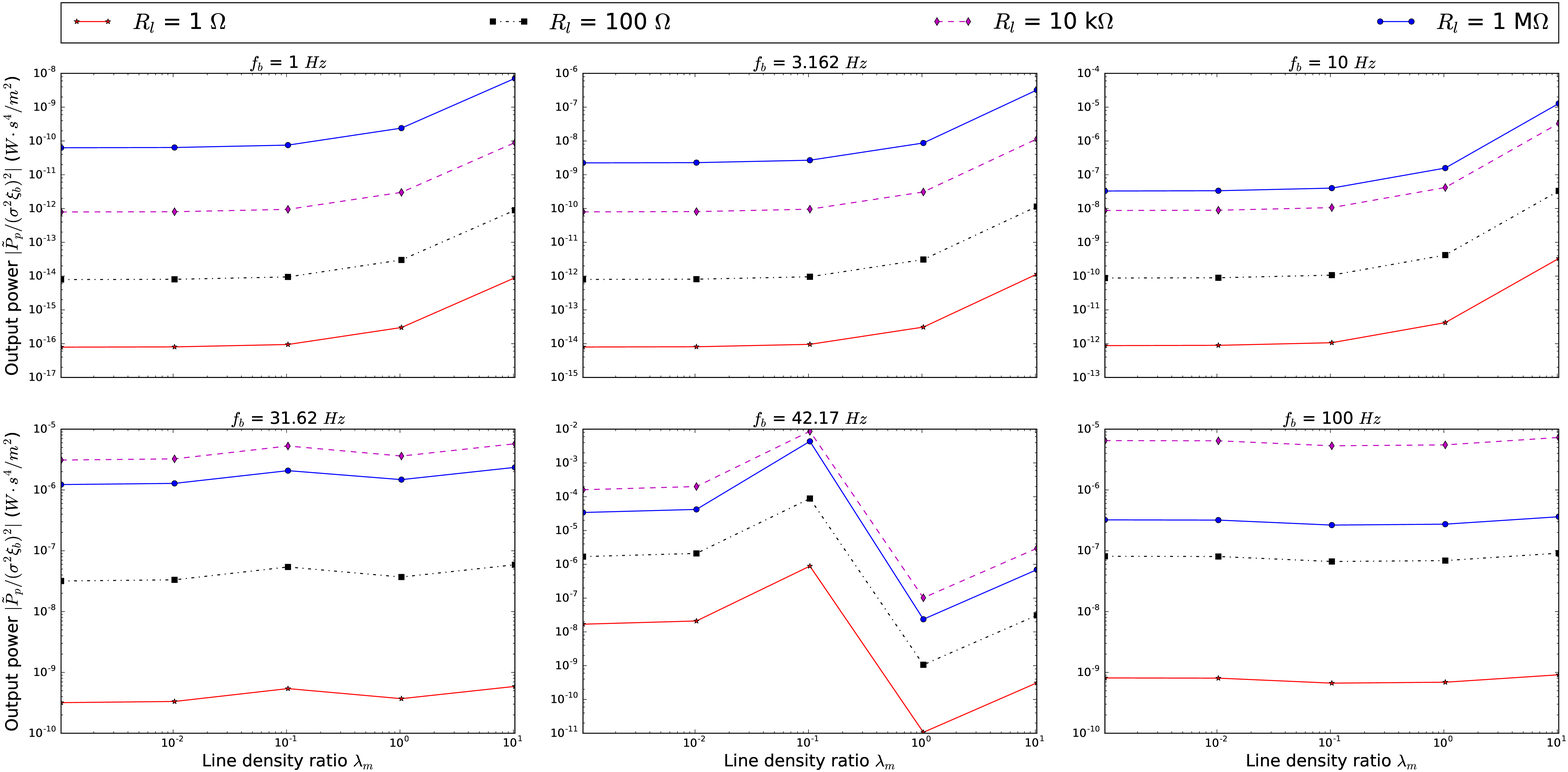}
    \caption{ Normalized output power of the PEHEE versus line density ratio $\lambda_m$ at different base excitation frequency $f_b$ and load resistance $R_l$. }
    \label{fig:fig_pow_fr_sl_Rl_sl_vs_lamm}
\end{figure}

A second plot is generated using the normalized output voltage $|\tilde{V}_p/(\sigma^2 \xi_b)|$ and normalized output power $|\tilde{P}_p/(\sigma^2 \xi_b)^2|$ versus line density ratio $\lambda_m$ at different values of load resistance $R_l$ and base excitation frequency $f_b$, as shown in Figure~\ref{fig:fig_vol_fr_sl_Rl_sl_vs_lamm} and Figure~\ref{fig:fig_pow_fr_sl_Rl_sl_vs_lamm}. 

It is obvious that for low base excitation frequency, like $f_b = 1 Hz$, normalized output voltage $|\tilde{V}_p/(\sigma^2 \xi_b)|$ and normalized output power $|\tilde{P}_p/(\sigma^2 \xi_b)^2|$ increase with respect to $\lambda_m$. When the base excitation frequency is increased towards the resonant frequency, say $f_b = 31.62 \ Hz$ as shown in Figure~\ref{fig:fig_vol_fr_sl_Rl_sl_vs_lamm} and Figure~\ref{fig:fig_pow_fr_sl_Rl_sl_vs_lamm}, the normalized output voltage $|\tilde{V}_p/(\sigma^2 \xi_b)|$ and normalized output power $|\tilde{P}_p/(\sigma^2 \xi_b)^2|$ change little with respect to $\lambda_m$, compared with the cases of smaller $\lambda_m$. For the case of $f_b = 41.27 \ Hz$, a significant change of $|\tilde{V}_p/(\sigma^2 \xi_b)|$ and $|\tilde{P}_p/(\sigma^2 \xi_b)^2|$ can be seen with respect to $\lambda_m$. While for the case of $f_b = 100 \ Hz$, the change becomes negligible again. That is to say, at the frequency around the resonant frequency in the considered range, the change of line density ratio $\lambda_m$ is helpful to improve the performance of the PEHEE. In the other cases, the change of $\lambda_m$ won't help too much.

\section{Conclusion and Discussion}

In this contribution, we propose the method of attaching elastic extensions to tube the performances of piezoelectric energy harvesters. Based on the Euler-Bernoulli's assumptions, we establish the electromechanical model of the proposed PEHEE and convert it into a boundary value problem. Systems parameters $\alpha$, $\lambda_l$, $\lambda_B$, and $\lambda_m$ as well as source vibration parameters $\beta$ and $\nu$ are isolated from the problem. A series of numerical simulations are conducted to explore the influences of different parameters. 

From the simulation results, several points may be summarized. Firstly, by changing the parameter values of $\lambda_l$, $\lambda_B$, and $\lambda_m$, we can introduce extra resonant modes in the considered frequency range. These resonant modes are potential for energy harvesting applications as the output level is usable. This provides a way to increase the working frequency range of practical piezoelectric energy harvesters. Secondly, by adding the elastic extensions, the performance of a CPEH can be tubed. When the interaction between the extension beam and primary beam is strong enough, it is possible to increase the bandwidth of the PEHEE. Besides, this kind of strong interaction also serves to increase the output performances of the PEHEE. Last but not the least important, the tuning effect of the elastic extension is the result of the combined actions of three parameters $\lambda_l$, $\lambda_B$, and $\lambda_m$. The increase of $\lambda_l$ and $\lambda_m$ tends to increase the number of introduced resonant modes, while the increase of $\lambda_B$ tends to decrease it.

In order to further increase the performance of piezoelectric energy harvesters, more researches are to be done in the future. Firstly, it seems promising by change the elastic extension into a full configurable elastic structure. Secondly, the asymptotic analysis of the elastic model may provide us a more detailed theoretical analysis of the proposed devices and guide our design. Thirdly, in the current analysis based on Euler-Bernoulli assumptions, the motion of the extension beam is actually of large amplitude. Thus it is necessary to take into account the geometric nonlinearity. What's more, the optimal parameter values for the proposed PEHEE need a more detailed analysis.

\section*{Acknowledgments}
The authors would like to thank the financial support from the National Natural Science Foundation of China (NSFC) under contract number 51705112 and 51905486. The research presented in this paper is funded by Open Foundation of the State Key Laboratory of Fluid Power and Mechatronic Systems (No. GZKF-2018017)

\section*{Reference}

\bibliography{maureenchou.bib}
\bibliographystyle{vancouver}

\end{document}